\documentclass[11pt,a4paper]{article}

\usepackage{jheppub}
\makeatletter
\gdef\@fpheader{Published in JHEP 06 (2026) 202; Erratum: JHEP 09 (2026) 149}
\makeatother

\usepackage{amsmath}
\usepackage{amssymb}
\usepackage{graphicx}
\usepackage{ascmac}
\usepackage{url}
\usepackage{comment}
\usepackage{physics}

\usepackage[justification=raggedright,singlelinecheck=false]{caption}
\usepackage{subcaption}

\usepackage{hyperref}
\hypersetup{%
 setpagesize=false,%
 bookmarks=true,%
 bookmarksdepth=tocdepth,%
 bookmarksnumbered=true,%
 colorlinks=true,%
 allcolors=blue,
 pdftitle={},%
 pdfsubject={},%
 pdfauthor={},%
 pdfkeywords={}}

\def\VEV#1{
\langle #1\rangle
}

\begin{document}
\title{Thermal Leptogenesis in $SO(10)\times U(1)_A$ SUSY GUT}

\author[a,b]{Nobuhiro Maekawa}
\author[a]{Kei Shibata}
\author[c]{Masato Yamanaka}

\affiliation[a]{Department of Physics,
Nagoya University, Nagoya 464-8602, Japan}
\affiliation[b]{Kobayashi-Maskawa Institute for the Origin of Particles and the
Universe, Nagoya University, Nagoya 464-8602, Japan}
\affiliation[c]{Department of Global Environment Studies, 
Hiroshima Institute of Technology, Hiroshima, 731-5193, Japan}

\emailAdd{maekawa@eken.phys.nagoya-u.ac.jp}
\emailAdd{shibata.kei.d6@s.mail.nagoya-u.ac.jp}
\emailAdd{m.yamanaka.km@cc.it-hiroshima.ac.jp}

\abstract{
We investigate thermal leptogenesis within a supersymmetric grand unified theory (SUSY GUT) based on the $SO(10) \times U(1)_A$ symmetry, where both the doublet--triplet splitting problem and the unrealistic Yukawa relations are resolved under the natural assumption that all symmetry-allowed interactions appear with $\mathcal{O}(1)$ coefficients. 
In this framework, the structures of the Dirac neutrino Yukawa couplings and right-handed neutrino masses are determined entirely by the symmetry. 
The baryon asymmetry of the Universe is computed taking into account the flavor effects, Higgs asymmetry contributions, and the impact of the second-lightest right-handed neutrino. 
While the predicted asymmetry is too small when all $\mathcal{O}(1)$ coefficients of the Dirac neutrino Yukawa couplings are set to unity, a moderate enhancement factor $r_1 \sim 4.4$ for the lightest right-handed neutrino mass reproduces the observed baryon asymmetry without spoiling low-energy neutrino data. 
This corresponds to a suppressed lightest neutrino mass, $m_{\nu_1} \sim (1/r_1) \times (\text{symmetry-determined value})$, typically $m_{\nu_1} \sim 5.5 \times 10^{-4}\,\text{eV}$. 
We further explore cases where the $\mathcal{O}(1)$ coefficients of the Dirac neutrino Yukawa couplings are $\pm 1$, and find that ten of the fifteen sign-flipped patterns successfully generate the observed baryon asymmetry for $r_1\leq 11$. 
Moreover, the others yield results of the correct order of magnitude, although they do not generate the observed Baryon asymmetry. 
These findings demonstrate that thermal leptogenesis is realized in this $SO(10)\times U(1)_A$ SUSY GUT, establishing a link between the observed baryon asymmetry and predictions for the lightest neutrino mass.
}

\arxivnumber{2511.11100}

\maketitle

\section{Introduction}
The Standard Model (SM), completed with the discovery of the Higgs boson in 2012, is a highly 
successful theory that explains most known phenomena.
Nevertheless, several phenomena remain beyond its scope, such as neutrino oscillations, the baryon asymmetry of the universe, dark matter, and inflation. 
Neutrino oscillations, however, can be accounted for by endowing neutrinos with tiny masses.  
The simplest and most natural way to generate neutrino masses is to introduce right-handed (RH) neutrinos in addition to the left-handed (LH) ones.
Since all quarks and leptons in the SM except neutrinos already possess both LH and RH components, the inclusion of RH neutrinos can be regarded as a natural extension of the SM.  
Moreover, introducing RH neutrinos is particularly compelling not only because it explains why LH neutrinos are much lighter than the other quarks and leptons~\cite{Minkowski:1977sc, Yanagida:1979as, Gell-Mann:1979vob, Mohapatra:1979ia}, but also because it can account for the baryon asymmetry of the universe via the leptogenesis mechanism~\cite{Fukugita:1986hr}.  

If we estimate the mass scale $M_R$ of RH neutrinos using the seesaw formula $m_\nu = y_{\nu_D}^2 v^2/M_R$, with the Dirac neutrino Yukawa coupling $y_{\nu_D}\sim 1$, Higgs vacuum expectation value $v\sim 175\,\mathrm{GeV}$, and neutrino mass $m_\nu\sim 0.05$ eV, we obtain $M_R=6\times 10^{14} \,{\rm GeV}:= \Lambda_R$, which is close to the supersymmetric grand unification scale $\Lambda_G=2\times 10^{16}\,\mathrm{GeV} $.  
This proximity strongly suggests a class of grand unified theories (GUTs) in which RH neutrinos acquire mass through the spontaneous breaking of the unified group.  
An $SO(10)$ GUT is precisely such a theory~\cite{Fritzsch:1974nn}: it unifies one generation of quarks and leptons (15 components) together with one RH neutrino (1 component) into a single 16-dimensional spinor representation, where the RH neutrino cannot acquire mass unless $SO(10)$ is broken.

Unfortunately, the simplest $SO(10)$ GUT suffers from a problem of Yukawa unification, 
$y_u=y_d=y_e=y_{\nu_D}$, arising from 
quark–lepton unification. Here, $y_u$, $y_d$, $y_e$, and $y_{\nu_D}$ are the Yukawa matrices of up-type 
quarks, down-type quarks, charged leptons, and Dirac neutrinos, respectively. The observed 
quark and lepton masses and mixing angles clearly contradict this unrealistic relation. 
Since the 1990s, when neutrino oscillations were first reported, $SO(10)$ GUTs have been 
extensively 
studied, including attempts to resolve this Yukawa unification problem.  

On the other hand, 
difficulties also arise in implementing thermal leptogenesis within $SO(10)$ GUTs, particularly when attempting to reproduce neutrino masses and large mixing angles.  
The observed neutrino masses and mixing angles suggest a weak mass hierarchy among LH neutrinos. 
In contrast, 
simple $SO(10)$ GUTs 
typically predict that
the Dirac neutrino Yukawa hierarchy 
mirrors 
that
of the up-type quarks, forcing the RH neutrino mass hierarchy to be the 
square of the up-type quark hierarchy: 
\[
M_{\nu_R}\sim \left(\left(\frac{m_u}{m_t}\right)^2,\; \left(\frac{m_c}{m_t}\right)^2,\;1\right)M_R
\sim (10^5 {\rm GeV},\, 10^{11} {\rm GeV},\, 10^{15} {\rm GeV}) .
\]  
however, 
requires the lightest RH neutrino mass to exceed $\sim 10^9\,\mathrm{GeV} $ in order to account for the observed baryon asymmetry\cite{Davidson:2002qv}. 
The above prediction clearly violates this Davidson–Ibarra bound.

One possible way to circumvent this problem is to incorporate the effects 
of the second-lightest RH neutrino~\cite{Pilaftsis:1998pd, Chun:2007ny, 
Nezri:2000pb, Akhmedov:2003dg, Asaka:2003fp, DiBari:2005st, Engelhard:2006yg, 
DiBari:2008mp, Kaladharan:2023zbr, DiBari:2025zlv, Vives:2005ra}.  
Usually, the lepton asymmetry produced by the second lightest RH neutrino is washed out by the lightest RH neutrino. However, if the washout effect is sufficiently weak, the effect of the second lightest RH neutrino is not negligible.
Moreover, at temperatures below $(1+\tan^2\beta)\times 10^{9}\,\mathrm{GeV}$~\cite{Davidson:2008bu}
\footnote{
Here, $\tan\beta := \langle H_u \rangle / \langle H_d \rangle$, since charged-lepton Yukawa couplings scale with $1/\cos\beta$, where
$H_u$ and $H_d$ denote the up- and down-type Higgs doublets, respectively. 
}, where the $\mu$ Yukawa  interaction becomes thermalized in addition to $\tau$ Yukawa interaction,  
each lepton flavor asymmetry must be treated separately
~\cite{Barbieri:1999ma, Endoh:2003mz, Nardi:2006fx, Abada:2006ea}. 
Note that the first two generation RH neutrino masses are below $(1+\tan^2\beta)\times 10^{9}\,\mathrm{GeV}$ in the above simple $SO(10)$ GUT where $\tan\beta\sim 50-60$. 
In such cases, the effects of the second-lightest RH neutrino can become more important~\cite{Vives:2005ra, Engelhard:2006yg, 
DiBari:2008mp, Kaladharan:2023zbr, DiBari:2025zlv}, because
the electron flavor asymmetry often is not washed out because of the smaller Dirac neutrino Yukawa coupling with electron. 
Since the mass of the second lightest RH neutrino satisfies the Davidson–Ibarra bound, it can potentially 
generate the observed baryon asymmetry, motivating many studies in this 
direction.  
Still, a fundamental question remains: why does the RH neutrino mass hierarchy 
become so much stronger than other Yukawa hierarchies?  
This poses a significant obstacle 
when 
one attempts 
to explain the Yukawa structure through flavor symmetries.

One of the simplest ways to relax the unrealistic Yukawa unification in 
$SO(10)$ GUTs is to introduce a $\mathbf{10}$-dimensional matter representation 
in addition to the three $\mathbf{16}$’s~\cite{Maekawa:2001uk, Maekawa:2001vt, 
Maekawa:2002mx}.  When the three spinors ${\bf 16}_i$ $(i=1,2,3)$ and one 
${\bf 10}$ are decomposed into $SU(5)$ multiplets, we obtain 
${\bf 16}_i={\bf 10}_i+{\bf \bar 5}_i+1_i$ and ${\bf 10}={\bf 5'}+{\bf \bar 5'}$, 
which yield four ${\bf \bar 5}$’s in total.  
Upon breaking $SO(10)$ to $SU(5)$, one pair of ${\bf 5}+{\bf \bar 5}$ becomes 
superheavy. Assuming that ${\bf \bar 5}_3$, which has larger Yukawa couplings, is the superheavy field, the SM fermions are then composed of ${\bf \bar 5}_1$, 
${\bf \bar 5'}$, and ${\bf \bar 5}_2$.
If
the first, second, and third generations of quarks and leptons are built from ${\bf \bar 5}_1$, ${\bf \bar 5'}$, and ${\bf \bar 5}_2$, respectively, the hierarchy among ${\bf \bar 5}$’s becomes very mild, almost flat. 
Consequently, compared with the ${\bf 10}_i$’s of $SU(5)$ that retain the 
original strong generational hierarchy, the ${\bf \bar 5}$ hierarchy is much 
weaker. This construction explains not only the differing mass hierarchies of 
quarks and leptons, but also why quark mixing angles are smaller than lepton 
mixing angles~\cite{Maekawa:2001uk, Maekawa:2001vt}.

The resulting Yukawa relations are $y_u \neq y_d = y_e^T \sim y_{\nu_D}^T$. 
If the $SU(5)$ relation $y_d = y_e^T$ is avoided, realistic Yukawa matrices 
can be
obtained. Importantly, realistic neutrino masses can also be realized with 
a much milder RH neutrino hierarchy, requiring only the first power of the 
up-quark hierarchy:  
\[
M_{\nu_R} = \left(\dfrac{m_u}{m_t},\, \dfrac{m_c}{m_t},\,1\right) 
(y_{\nu_D})_{33} M_R
\simeq (10^8 {\rm GeV},\, 10^{11} {\rm GeV},\, 10^{13} {\rm GeV}) ,
\]  
where $(y_{\nu_D})_{33}$, the (33) element of the Dirac neutrino Yukawa 
matrix, is significantly smaller than unity (taken here as $0.04$) because 
${\bf \bar 5}_3$ became superheavy.

Notably, this framework naturally reproduces the various quark and lepton mass 
hierarchies and mixing angles within $SO(10)$ GUTs when combined with family 
symmetries such as anomalous $U(1)_A$.  
Nevertheless, even in this case, the lightest RH neutrino mass is 
$10^8\,\mathrm{GeV}$ an improvement, but still below the Davidson–Ibarra 
bound.  
It has therefore been argued that non-thermal leptogenesis mechanisms may be required to reproduce the observed baryon asymmetry~\cite{Asaka:2003fp}.

In this paper, we numerically calculate the baryon asymmetry via thermal 
leptogenesis, taking into account supersymmetric effects, flavor effects, 
and the role of the second-lightest RH neutrino in a natural $SO(10)$ GUT, 
where the above Yukawa structures are realized but the $SU(5)$ relations are avoided.  
In this GUT scenario, the doublet–triplet splitting problem, one of the most serious issues in SUSY GUTs, can also be solved under natural assumptions.  
We do not consider the constraint on the reheating temperature from the gravitino problem~\cite{Weinberg:1982zq, Khlopov:1984pf, Ellis:1984eq, Reno:1987qw, Kawasaki:1994af, Kohri:2001jx, Kawasaki:2004qu, Kawasaki:2008qe}, in which thermally produced gravitinos spoil the success of the big-bang nucleosynthesis (BBN). This is because the problem strongly depends on the mass spectrum of SUSY particles, which is independent of the idea of GUT.  For example, it is known that heavy gravitino with a mass larger than $\mathcal{O}(100)\,\mathrm{TeV}$
can decay before the BBN starts~\cite{Luty:2001jh, Luty:2001zv, Ibe:2004tg}.  

The baryon asymmetry computed in this model is much smaller than the observed value when all $\mathcal{O}(1)$ coefficients are taken as unity.  
However, increasing the $\mathcal{O}(1)$ coefficient $r_1$ associated with the lightest RH neutrino mass yields a sufficiently large baryon asymmetry.  
Note that raising the lightest RH neutrino mass does not affect neutrino observables, except for the lightest active neutrino mass, which has not been unmeasured yet. 
We therefore analyze how large the lightest RH neutrino mass must be to 
reproduce the observed baryon asymmetry.  
We find 
that for $r_1\simeq 4.4$, the observed baryon asymmetry can indeed be obtained.  
In this case, the mass of the lightest LH neutrino is predicted to be $1/r_1$ 
of the symmetry-determined value.

For comparison, previous numerical study~\cite{Ishihara:2015uua} considering supersymmetric and flavor effects have been performed in $E_6$ GUTs, where about 16 times larger lightest RH neutrino mass than the value fixed by the symmetry was required.  
In $E_6$ GUTs, the presence of six RH neutrinos allows for a separation between 
those responsible for LH neutrino masses and those relevant for leptogenesis.  
Thus, even when the lightest RH neutrino mass is increased, no predictions 
arise for low-energy observables. Furthermore, those studies did not include 
the effects of the second-lightest RH neutrino.

The structure of this paper is as follows.  
In section~\ref{Sec:naturalSO10GUT}, we review the natural $SO(10)$ GUT.  
Section~\ref{Sec:naturalSO10Lepto} provides a review of leptogenesis in the natural $SO(10)$ GUT, 
while section~\ref{Sec:NumericalRes} presents our numerical results.  
In Section~\ref{Sec:Discussion} we compare our results in natural $SO(10)$ GUT with those in natural $E_6$ GUT.
Finally Sec.~\ref{Sec:Conclusion} is devoted to the summary.

\section{Natural $\mathbf{SO(10)}$ GUT}
\label{Sec:naturalSO10GUT}
In this section, we briefly review the natural $SO(10)$ GUT\cite{Maekawa:2001uk,Maekawa:2001vt,Maekawa:2002mx}. Table~\ref{SO10} lists typical quantum numbers for a natural $SO(10)$ GUT.
In simple terms, a “natural” GUT is one in which the two issues present in SUSY GUTs —the doublet–triplet splitting problem and the unrealistic Yukawa relations arising from matter unification—are resolved under the natural assumption that all interactions allowed by the symmetry are present with $\mathcal{O}(1)$ coefficients. 
An essential role is played by an anomalous $U(1)_A$ gauge symmetry.
Although anomalous at first sight, the $U(1)_A$ anomalies are canceled by the Green–Schwarz mechanism\cite{Green:1984sg}, i.e., by a dilaton shift. 
One may regard this as a $U(1)$ gauge symmetry with a Fayet–Iliopoulos (FI) term,
\[
\xi^2 \int d^2\theta\,d^2\overline{\theta} V_A,
\]
where $\xi$ is the FI parameter and $V_A$ is the vector supermultiplet of $U(1)_A$.

\begin{table}[t]
\caption{Field content of a natural $SO(10)$ GUT with $U(1)_A$ charges which are denoted by small characters. The $\pm$ labels denote the $Z_2$ parity.}
  \begin{center}
    \begin{tabular}{|c||c|c|c|}
      \hline
      $SO(10)$ & negatively charged fields & positively charged fields & matter fields \\ \hline \hline
      {\bf 45}  & $A\,(a=-1,-)$ & $A'\,(a'=3,-)$ & \\ \hline
      {\bf 16} & $C\,(c=-4,+)$ & $C'\,(c'=3,-)$  & $\Psi_i\bigl(\psi_1=\frac{9}{2},\, \psi_2=\frac{7}{2},\, \psi_3=\frac{3}{2},\, +\bigr)$ \\ \hline
      ${\bf \overline{16}}$ & $\bar C\,(\bar c=-1,+)$ & $\bar C'\,(\bar c'=7,-)$ & \\ \hline
      {\bf 10} & $H\,(h=-3,+)$  & $H'\,(h'=4,-)$ & $T\,(t=\frac{5}{2},+)$ \\ \hline
      1 &  $\begin{array}{c}\Theta\,(\theta=-1,+)\\ Z\,(z=-2,-),\ \bar Z\,(\bar z=-2,-) \end{array}$ & $S\,(s=5,+)$ &
      \\ \hline
    \end{tabular}
    \label{SO10}
  \end{center}
\end{table}

In many constructions, the anomalous $U(1)_A$ also serves as a flavor symmetry. 
For example, retaining only the matter fields $\Psi_i$ $(i=1,2,3)$, the Higgs field $H$, and the flavon $\Theta$, the $U(1)_A$–invariant superpotential is
\begin{equation}
W_Y \;=\; y_{ij}\!\left(\frac{\Theta}{\Lambda}\right)^{\psi_i+\psi_j+h}\!\Psi_i\Psi_j H,
\end{equation}
where $\Lambda$ is the cutoff scale, $y_{ij}$ are $\mathcal{O}(1)$ coefficients, and $\psi_i$ and $h$ are the $U(1)_A$ charges of $\Psi_i$ and $H$, respectively. 
We take the $U(1)_A$ charge of the flavon $\Theta$ to be $-1$. 
Imposing $D$–flatness fixes the VEV as $\VEV{\Theta}=\xi=\lambda\Lambda$, and in this paper we use $\lambda\simeq 0.22$.  
Thus,
\begin{equation}
W_Y \;=\; y_{ij}\,\lambda^{\psi_i+\psi_j+h}\,\Psi_i\Psi_jH.
\end{equation}
The crucial point is that, under the natural assumption that all symmetry-allowed operators —including higher-dimensional ones— appear with $\mathcal{O}(1)$ coefficients, hierarchical Yukawa structures are reproduced automatically\cite{Ibanez:1991hv}. 
When this assumption is extended to the whole theory, including the Higgs sector responsible for breaking the unified gauge group, the resulting framework is referred to as a natural GUT. 
Apart from the $\mathcal{O}(1)$ coefficients, the theory is then essentially fixed by symmetry. 

The VEV of a field $F$ is determined by its $U(1)_A$ charge $f$ as\cite{Maekawa:2001uk,Maekawa:2001vt,Maekawa:2002mx}
\begin{equation}
\label{VEV}
\VEV{F}\sim \begin{cases}
\lambda^{-f}\Lambda & (f\leq 0),\\[2pt]
0 & (f>0).
\end{cases}
\end{equation}
For example, the positively charged fields $A'$, $C'$, $\bar C'$, $H'$, and $S$ in Table~\ref{SO10} have vanishing VEVs, while the negatively charged fields
$A$, $C$, $\bar C$, $\Theta$, $Z$, and $\bar Z$ can develop nonzero VEVs.
The VEV $\VEV{A}\sim \lambda^{-a}\Lambda$ breaks $SO(10)\to SU(3)_C\times SU(2)_L\times SU(2)_R\times U(1)_{B-L}$, and the VEVs $\VEV{C}=\VEV{\bar C}\sim \lambda^{-\frac{(c+\bar c)}{2}}\Lambda$
\footnote{
The gauge-invariant VEV $\VEV{\bar C C}$ scales as $\lambda^{-(c+\bar c)}\Lambda^2$. 
$SO(10)$ $D$–flatness requires $\VEV{C}=\VEV{\bar C}$, hence $\VEV{C}=\VEV{\bar C}\sim \lambda^{-\frac{c+\bar c}{2}}\Lambda$.
}
break $SU(2)_R\times U(1)_{B-L}\to U(1)_Y$.

As discussed in Ref.~\cite{Maekawa:2001vt,Maekawa:2002mx}, gauge coupling unification in a natural GUT requires the cutoff $\Lambda$ to coincide with the usual GUT scale $\Lambda_G\sim 2\times 10^{16}\,\mathrm{GeV}$. 
An interesting prediction is enhanced proton decay, since the adjoint VEV $\VEV{A}\sim \lambda^{-a}\Lambda_G$ is smaller than $\Lambda_G$ itself.
It is particularly important that positively charged fields have vanishing VEVs
\footnote{
Strictly speaking, we assume all positively charged fields have vanishing VEVs. 
Because $U(1)_A$ $D$–flatness imposes $\xi^2+\sum z_i|Z_i|^2=0$ with $\xi^2\ll \Lambda^2$, negatively charged fields are forced to acquire small VEVs. Such vacua can have interesting low-energy implications.
}
; consequently, operators with negative total $U(1)_A$ charge are forbidden by symmetry. 
This “supersymmetric zero mechanism” is crucial for resolving the doublet–triplet splitting problem.

Under this natural assumption, both issues of SUSY GUTs can be addressed. 
For instance, since $\VEV{A}\sim\lambda^{-a}\Lambda$, the contribution from the higher-dimensional operator\\
$\lambda^{\psi_i+\psi_j+a+h} \Psi_i (A/\Lambda) \Psi_j H$
is of the same order as that from
$\lambda^{\psi_i+\psi_j+h}\,\Psi_i\,\Psi_j\,H.$
Thus, the unrealistic Yukawa relations from matter unification are naturally avoided. 
We do not elaborate on the solution to the doublet–triplet splitting problem here, as it is not directly related to the subject of this paper.

\subsection{Quark and Lepton Mass Matrices}
For leptogenesis in the natural GUT, it is important that both the Dirac neutrino Yukawa couplings and the RH-neutrino masses are symmetry-determined. 
Their explicit values are summarized in Table~\ref{parameters}.
\begin{table}[t]
  \centering
  \caption{
Three RH-neutrino masses and Dirac neutrino Yukawa entries fixed by symmetry in this scenario, up to $\mathcal{O}(1)$ coefficients.
}
\begin{tabular}{cll}
  parameter && value\\
  \hline\hline
  Wolfenstein parameter & $\lambda$ & $0.22$\\
  GUT scale & $\Lambda_G$ & $2.0\times 10^{16}\,\mathrm{GeV}$\\
  $N_1$ mass & $\overline{M}_1=\lambda^{12}\Lambda_G$ & $2.6\times 10^{8}\,\mathrm{GeV}$\\
  $N_2$ mass & $M_2=\lambda^{10}\Lambda_G$ & $5.3\times 10^{9}\,\mathrm{GeV}$\\
  $N_3$ mass & $M_3=\lambda^{6}\Lambda_G$ & $2.7\times 10^{12}\,\mathrm{GeV}$\\
  $(e1)$ of $(y_{\nu_D})_{\alpha i}$ &
  $(y_{\nu_D})_{e1}=\lambda^{6}$ & $1.1\times 10^{-4}$\\
  $(\mu 1)$ of $(y_{\nu_D})_{\alpha i}$ &
  $(y_{\nu_D})_{\mu1}=\lambda^{5.5}$ & $2.4\times 10^{-4}$\\
  $(\tau 1)$ of $(y_{\nu_D})_{\alpha i}$ &
  $(y_{\nu_D})_{\tau1}=\lambda^{5}$ & $5.2\times 10^{-4}$\\
  $(e2)$ of $(y_{\nu_D})_{\alpha i}$ &
  $(y_{\nu_D})_{e2}=\lambda^{5}$ & $5.2\times 10^{-4}$\\
  $(\mu 2)$ of $(y_{\nu_D})_{\alpha i}$ &
  $(y_{\nu_D})_{\mu2}=\lambda^{4.5}$ & $1.1\times 10^{-3}$\\
  $(\tau 2)$ of $(y_{\nu_D})_{\alpha i}$ &
  $(y_{\nu_D})_{\tau2}=\lambda^{4}$ & $2.3\times 10^{-3}$\\
  $(e3)$ of $(y_{\nu_D})_{\alpha i}$ &
  $(y_{\nu_D})_{e3}=\lambda^{3}$ & $1.1\times 10^{-2}$\\
  $(\mu 3)$ of $(y_{\nu_D})_{\alpha i}$ &
  $(y_{\nu_D})_{\mu3}=\lambda^{2.5}$ & $2.3\times 10^{-2}$\\
  $(\tau 3)$ of $(y_{\nu_D})_{\alpha i}$ &
  $(y_{\nu_D})_{\tau3}=\lambda^{2}$ & $4.8\times 10^{-2}$\\
  \hline
 \end{tabular}  
 \label{parameters}
\end{table}

Note that the Dirac neutrino Yukawas are not as large as those predicted by the naive $SO(10)$ relation $y_u=y_{\nu_D}$. 
This is because, in addition to the three spinor matter fields $\Psi_i$ (each in the {\bf 16}, $i=1,2,3$), one introduces a vector $T$ (in the {\bf 10}). 
Upon decomposition under $SU(5)$,
\begin{eqnarray}
{\bf 16}_i &=& {\bf 10}_i +{\bf \bar 5}_i+ 1_i, \\
{\bf 10}_T &=&{\bf 5}_T +{\bf \bar 5}_T,
\end{eqnarray}
so that four ${\bf \bar 5}$ representations appear in total. 
One of these, together with ${\bf 5}_T$, becomes superheavy after developing the VEVs $\langle C\rangle=\langle \bar C\rangle$ which break $SO(10)$ into $SU(5)$, and the SM fermions correspond to the remaining three ${\bf \bar 5}$’s.
It is reasonable that the ${\bf \bar 5}_3$ with the largest Yukawas becomes the heavy mode ${\bf \bar 5}_M$, so that the three generations are mainly
\begin{equation}
   ({\bf \bar 5}_{1}^0,\, {\bf \bar 5}_{2}^0,\, {\bf \bar 5}_{3}^0) \;\sim\; ({\bf \bar 5}_{1},\, {\bf \bar 5}_{T},\, {\bf \bar 5}_{2}).
\end{equation} 
This result explains not only the differing mass hierarchies of quarks and leptons but also why quark mixing angles are smaller than lepton mixing angles, that will be shown below.

First of all, the Dirac neutrino Yukawas are suppressed. Indeed, since
\begin{equation}
{\bf \bar 5}_3\sim {\bf \bar 5}_M + \lambda^\Delta\,{\bf \bar 5}_{2}^0 + \cdots,
\end{equation}
the state ${\bf \bar 5}_{2}^0$ inherits Yukawa couplings. 
In the model of Table~\ref{SO10}, one finds $\Delta= t-\psi_3+\tfrac{1}{2}(\bar c-c)=2.5$, leading to
\begin{equation}
y_u\sim \begin{pmatrix}
\lambda^6 & \lambda^5 & \lambda^3 \\
\lambda^5 & \lambda^4 & \lambda^2 \\
\lambda^3 & \lambda^2 & 1
\end{pmatrix},\qquad
y_d\sim y_e^{T}\sim y_{\nu_D}^{T}\sim \begin{pmatrix}
\lambda^6 & \lambda^{5.5} & \lambda^{5} \\
\lambda^{5} & \lambda^{4.5} & \lambda^{4} \\
\lambda^{3} & \lambda^{2.5} & \lambda^{2}
\end{pmatrix},
\end{equation}
which yield realistic hierarchies. 
The $SU(5)$ relation $y_d = y_e^{T}$ can be avoided by contributions involving the adjoint VEV $\VEV{A}\sim \lambda^{-a}\Lambda$, because the higher-dimensional operator $\lambda^{\psi_i+\psi_j+a+h}\Psi_i A \Psi_j H$ is of the same order as $\lambda^{\psi_i+\psi_j+h}\Psi_i\Psi_jH$ and does not respect exact $SU(5)$ relations. 
Consequently, realistic Yukawa matrices are obtained when $\tan\beta\sim 7-8$.

The RH-neutrino mass arises from
\begin{equation}
\lambda^{\psi_i+\psi_j+2\bar c}\,\Psi_i\Psi_j\,\bar C^{\,2}
\;\Rightarrow\;
(M_{\nu_R})_{ij}\sim \lambda^{\psi_i+\psi_j+\bar c-c}\,\Lambda,
\end{equation}
where $\VEV{\bar C}=\VEV{C}=\lambda^{-\frac{1}{2}(c+\bar c)}\Lambda$ is fixed by the charges. 
Without loss of generality, we may take $(M_{\nu_R})$ diagonal, $(M_{\nu_R})_{ij}=\lambda^{\psi_i+\psi_j+\bar c-c}\Lambda\,\delta_{ij}$. 
Then the LH-neutrino masses are generated via the seesaw:
\begin{equation}
(M_\nu)\;=\; (M_{\nu_D})\,(M_{\nu_R})^{-1}\,(M_{\nu_D})^{T}
\;\sim\; \lambda^{4+h+c-\bar c}
\begin{pmatrix}
\lambda^{2} & \lambda^{1.5} & \lambda \\
\lambda^{1.5} & \lambda & \lambda^{0.5} \\
\lambda & \lambda^{0.5} & 1
\end{pmatrix}\frac{v^2}{\Lambda},
\label{eq:nuYukawa}
\end{equation}
with $M_{\nu_D}:= y_{\nu_D}v$ and $v\sim 175$ GeV the SM Higgs VEV, yielding the light neutrino masses as
$(m_{\nu_1},m_{\nu_2}, m_{\nu_3})\sim (0.0015,0.007,0.03)$ eV except the $O(1)$ coefficients, which roughly explain the observed neutrino mass differnces
$|m_{\nu_3}^2-m_{\nu_2}^2|\sim 2.5\times 10^{-3} (\rm eV)^2$ and $m_{\nu_2}^2-m_{\nu_1}^2\sim 7.5\times 10^{-5} (\rm eV)^2$.
The quark and lepton mixing matrices are
\begin{equation}
V_{CKM}\sim \begin{pmatrix}
1 & \lambda & \lambda^{3} \\
\lambda & 1 & \lambda^{2} \\
\lambda^{3} & \lambda^{2} & 1
\end{pmatrix},\qquad
V_{MNS}\sim \begin{pmatrix}
1 & \lambda^{0.5} & \lambda \\
\lambda^{0.5} & 1 & \lambda^{0.5} \\
\lambda & \lambda^{0.5} & 1
\end{pmatrix}.
\end{equation}
It is obvious that this model predicts normal hierarchy and
$V_{e2}\sim V_{\mu 3}\sim 0.5$ and $V_{e3}\sim 0.22$, which explain the observed neutrino data $\sin^2(\theta_{12})\sim 0.3$, $\sin^2(\theta_{23})\sim 0.5$, and $\sin^2(\theta_{13})\sim 0.02$.

It is important that $M_{\nu}$ is also fixed by the $U(1)_A$ charges. 
For example,
\begin{equation}
(M_{\nu})_{33} \;=\; \lambda^{\,2\psi_{2}+2h-\bar c+c}\,\frac{v^2}{\Lambda},
\end{equation}
so that, even after summing over RH generations, the dependence on the RH charge $\psi_i$ cancels. 
This indicates that when RH-neutrino masses are symmetry-determined, the individual RH contributions to LH masses are of comparable order.
Therefore, even if $(M_{\nu_R})_{11}$ accidentally becomes larger than the symmetry-determined value, the order of each component of $M_\nu$ does not change because the contributions to the $M_\nu$ from the other RH neutrinos do not change although the contribution from the lightest RH neutrino is suppressed. That means that the predicted orders of the observed neutrino data,  $|m_{\nu_3}^2-m_{\nu_2}^2|$, $m_{\nu_2}^2-m_{\nu_1}^2$ and the lepton mixings do not change as long as the enhanced $(M_{\nu_R})_{11}$ is smaller than $M_2$.
For later use, we parametrize an $\mathcal{O}(1)$ coefficient $r_1$ for the lightest RH-neutrino mass, $M_1:=r_1\,\overline{M}_1$, since the generated baryon asymmetry is highly sensitive to this parameter (all other $\mathcal{O}(1)$ coefficients are set to unity). 
Here $\overline{M}_1$ denotes the mass for $r_1=1$. 
Since we assume that only $(M_{\nu_R})_{11}$ becomes $r_1$ times larger than the symmetry-determined value, 
the observed neutrino masses and mixings remain unchanged even for $1\ll r_1< M_2/\overline{M_1}\sim 20$.
Because the three RH states contribute comparably to $M_\nu$, the heavier two LH eigenvalues are essentially unchanged. 
However, using
\begin{align}
m_{\nu_1}m_{\nu_2}m_{\nu_3}=\det(M_\nu)=\frac{\bigl[\det(M_{\nu_D})\bigr]^2}{\det(M_{\nu_R})}
=\frac{\bigl[\det(M_{\nu_D})\bigr]^2}{M_1 M_2 M_3}, \label{detmM}
\end{align}
we see that the lightest LH-neutrino mass is reduced by a factor $1/r_1$ relative to its symmetry value. 
Since the mixing angles are unaffected, increasing $M_1$ by $r_1$ lowers $m_{\nu 1}$ by $1/r_1$, while other low-energy neutrino observables remain essentially unchanged.

Here, we just mention that $r_1$ is naturally expected to be larger than 1 because we have a lot of independent operators which
contribute to $(M_R)_{11}$. For example, an operator $\lambda^{2\psi_1+2{\bar c}+2a}\Psi_1^2A^2{\bar C}^2$ becomes
$\lambda^{2\psi_1+2{\bar c}}\Psi_1^2{\bar C}^2$ after developing the VEV $\VEV{A}\sim \lambda^{-a}$, which gives the same order of contributions to $(M_R)_{11}$ as the original operator $\lambda^{2\psi_1+2{\bar c}}\Psi_1^2{\bar C}^2$. 
Of course, similar effect appears in other operators. It is obvious that the operator with larger total $U(1)_A$ charge has larger number of the independent operators which give the comparable contribution.  Since the total $U(1)_A$ charge of the operator $\Psi_1^2{\bar C}^2$, which contributes to $(M_R)_{11}$ after developing the VEV of $\bar C$, 
is larger than the $U(1)_A$ charges of the other operators which contribute to Yukawa interactions,   this effect works stronger for $(M_R)_{11}$ than the Yukawa couplings.
Therefore, the $O(1)$ coefficient $r_1$ is naturally expected to be larger than 1. (The details are discussed in Ref. ~\cite{Ishihara:2015uua}.)
\section{
Leptogenesis in $\mathbf{SO(10) \times U(1)_A}$ model 
}
\label{Sec:naturalSO10Lepto}

A $B-L$ asymmetry is dynamically generated through CP-violating (inverse) decays and scatterings of RH neutrinos. The baryon ($B$) and lepton ($L$) asymmetries are defined as the net differences between baryons and antibaryons, and between leptons and antileptons, respectively. 
Part of the generated $B-L$ asymmetry is converted into a baryon asymmetry via sphaleron processes~\cite{Harvey:1990qw}, 
\begin{align}
    Y_B(t_0) = \frac{8}{23} Y_{B-L}(t_f), 
\end{align}
where $t_f$ denotes the freeze-out time of $Y_{B-L}$ and $t_0$ the present epoch. 
To account for the observed baryon asymmetry, one requires $Y_{B-L}^{\text{obs}} \simeq 2.5 \times 10^{-10}$~\cite{Planck:2018vyg}. 
The final freeze-out value $Y_{B-L}(t_f)$ is obtained by solving a coupled network of Boltzmann equations (see Appendix A). 
The generated $B-L$ asymmetry depends on both the neutrino Yukawa couplings and the RH-neutrino mass spectrum, which are dictated by the $SO(10)\times U(1)_A$ GUT structure. Thus, the prediction for baryon asymmetry directly reflects the underlying GUT model. 
To highlight this connection with the model parameters, we first outline the basic mechanism of unflavored (``vanilla’’) leptogenesis, and then discuss the flavor-dependent dynamics essential for evaluating $B-L$ asymmetry within $SO(10)\times U(1)_A$.

At temperature $T \gtrsim (1+\tan^2\beta)\times 10^{12}$~GeV, charged-lepton Yukawa interactions are out of thermal equilibrium. 
If $M_1>(1+\tan^2\beta)\times 10^{12}$~GeV, the generated lepton asymmetry is mainly controlled by the CP-asymmetry parameters $\epsilon_i$ and decay parameters $K_i$. 
The total CP-asymmetry can be expressed as~\cite{Covi:1996wh}
\begin{align}
    \epsilon_i
    :=&
    \epsilon^{\ell_L}_{N_i}
    +\epsilon^{\widetilde{\ell}_L}_{N_i} 
    +\epsilon^{\ell_L}_{\widetilde{N}_i}
    +\epsilon^{\widetilde{\ell}_L}_{\widetilde{N}_i}
    =
    4 \epsilon^{\ell_L}_{N_i}\nonumber
    \\ \simeq &
    -\frac{1}{2\pi} \frac{1}{(y_{\nu_D}^\dag y_{\nu_D})_{ii}}
    \sum_{j}^3 \Im \left[ (y_{\nu_D}^\dag y_{\nu_D})^2_{ij} \right]
    f(M_j^2/M_i^2),
\end{align}
where
\begin{align}  
    \epsilon^{\ell_L}_{N_i}
    :=&
    \frac{\Gamma(N_i\rightarrow \ell_L+H_u)-\Gamma(N_i\rightarrow \overline{\ell}_L+H_u^\dag)}
    {\Gamma(N_i\rightarrow \ell_L+H_u)+\Gamma(N_i\rightarrow \overline{\ell}_L+H_u^\dag)}
    \nonumber 
    \\ \simeq& 
    -\frac{1}{8\pi}\frac{1}{(y_{\nu_D}^\dag y_{\nu_D})_{ii}} 
    \sum_{j}^3 \Im \left[ (y_{\nu_D}^\dag y_{\nu_D})^2_{ij} \right]
    f(M_j^2/M_i^2), 
    \\&
    f(x) = 
    \sqrt{x} \left( \frac{2}{x-1}+\ln\left[\frac{1+x}{x}\right] \right).  
\end{align} 
The other CP asymmetries $\epsilon^{\widetilde{\ell}_L}_{N_i}$, $\epsilon^{\ell_L}_{\widetilde{N}_i}$, and $\epsilon^{\widetilde{\ell}_L}_{\widetilde{N}_i}$ are defined analogously. 
Here $\ell_L$, $\widetilde{\ell}_L$, and $\widetilde{N}_i$ denote lepton doublets, slepton doublets, and RH sneutrinos, respectively. The resulting $B-L$ asymmetry is essentially proportional to $\epsilon_i$ (up to washout effects).
The efficiency of washout is governed by the decay parameter
\begin{align}   
    K_i :=
    \frac{\Gamma_{N_i}(T=0)}{H(T=M_i)} 
    \simeq 
    \sqrt{\frac{45}{4\pi^3g_\ast}} 
    \frac{\left(y_{\nu_D}^\dag y_{\nu_D}\right)_{ii}}{8\pi} 
    \frac{M_{pl}}{M_i},
\end{align}
where $g_\ast$ counts the effective number of relativistic degrees of freedom. 

In the literature~\cite{Buchmuller:2004nz, Blanchet:2006be}, if $K_1\lesssim 3$, the thermally produced $B-L$ asymmetry is dependent on the initial conditions. Thus, the contribution from $N_2$ must be taken into account, since the initial conditions are determined by the decay of $N_2$.
For $K_1 \gtrsim 3$, washout processes erase any dependence on initial conditions and suppress contributions from heavier RH neutrinos. 
In contrast, for $K_1 \lesssim 3$, washout is weak, allowing asymmetries generated by heavier states to survive.

When the temperature falls below $(1+\tan^2\beta)\times 10^{12}$~GeV, the $\tau$ Yukawa interaction equilibrates, marking the onset of the flavored regime of leptogenesis when $M_1<(1+\tan^2\beta)\times 10^{12}$~GeV. 
As temperature decreases further below $(1+\tan^2\beta)\times 10^{9}$~GeV, $\mu$ Yukawa interaction 
equilibrates, causing the lepton asymmetry to decohere into independent three flavor components when $M_1<(1+\tan^2\beta)\times 10^{9}$~GeV. 
Accounting for these flavor effects is crucial, as they can qualitatively change the efficiency and even the sign of the final baryon asymmetry.
This phenomenon is known as the flavor effect in leptogenesis 
~\cite{Barbieri:1999ma, Endoh:2003mz, Nardi:2006fx, Davidson:2008bu, Abada:2006ea}.

In this paper, we track the asymmetries generated by $N_1$ and $N_2$, while neglecting $N_3$ because of its much higher mass scale ($M_3 \gg M_{1,2}$) and the strong washout expected from $K_2 \gg 1$. We focus on the interplay between three flavored CP asymmetries $\epsilon_{\alpha i}$ and flavored decay parameters $K_{\alpha i}\,(\alpha=e,\mu,\tau)$ because the $SO(10)\times U(1)_A$ GUT has $\overline{M}_1 \sim 2.6\times 10^8$\,GeV and $M_2 \sim 5.3\times 10^9\,\mathrm{GeV}$, and $\tan\beta\sim 7-8$.
Therefore, we assume that the reheating temperature is larger than $M_2$. If we take into account of the sufficient thermalization, we require the reheating temperature higher than about $10\times M_2$.
Numerically, 
we place the thermal window in the range $10^7 \lesssim T \lesssim 10^{11}\,\mathrm{GeV}$, where the lower bound is determined by $M_1$.
In this regime, the $\tau$ and $\mu$ Yukawa interactions are in thermal equilibrium, whereas the electron Yukawa is not. 

Earlier studies have shown that flavor effects can significantly alter the final baryon asymmetry, especially in the strong washout regime ($K \gg 1$)~\cite{Nardi:2006fx, Barbieri:1999ma}. 
The enhancement is particularly pronounced for hierarchical neutrino Yukawa couplings, which is exactly the case in the $SO(10)\times U(1)_A$ models where most $\mathcal{O}(1)$ coefficients are taken to be unity except for the one controlling the mass of the lightest RH neutrino. 
In this setting, the flavored decay parameters, $K_{\alpha i} := \Gamma_{N_i \to \ell_\alpha H}(T=0)/H(T=M_i)$, read 
\begin{equation}
\begin{split}
    K_{\alpha 1} = \overline{K}_{\alpha} 
    \frac{\overline{M}_1}{M_1}
    = \frac{\overline{K}_{\alpha}}{r_1}, ~~~ 
    K_{\alpha 2} = \overline{K}_{\alpha}
\label{Eq:K_alphai}	
\end{split}     
\end{equation}
with $\overline{K}_e=1.9$, $\overline{K}_\mu=8.7$, and $\overline{K}_\tau=40$, where $\overline{K}_\alpha\equiv K_{\alpha 1}|_{M_1=\overline{M_1}}$.  Note that $K_{\alpha 2}=K_{\alpha 3}=\overline{K_\alpha}$.
These values confirm that the $\tau$ flavor asymmetry is strongly washed out, while the $e$ and $\mu$ channels are comparatively weaker. Since $r_1$ parameterizes the variation in the lightest RH neutrino mass, the washout of $N_1$ is directly governed by this coefficient. 
Thus, the $SO(10)\times U(1)_A$ structure tightly correlates flavor hierarchies with the flavored lepton asymmetries.

\begin{table}[t]
\centering
\caption{Flavored CP asymmetries and decay parameters for $r_1=1$ 
in $SO(10)\times U(1)_A$ GUT scenario.}
\begin{tabular}{c||c|c|c|c|c|c|c|c}
    &$\epsilon_i$&$\epsilon_{e i}$&$\epsilon_{\mu i}$&$\epsilon_{\tau i}$&$K_i$&$K_{ei}$&$K_{\mu i}$&$K_{\tau i}$\\
    \hline
    \hline
    $N_1$&$-1.6\times 10^{-7}$&$-6.1\times 10^{-9}$&$-2.8\times 10^{-8}$&$-1.3\times 10^{-7}$&$51$&$1.9$&$8.7$&$40$\\
    \hline
    $N_2$&$-1.7\times 10^{-6}$&$-6.1\times 10^{-8}$&$-2.8\times 10^{-7}$&$-1.3\times 10^{-6}$&$51$&$1.9$&$8.7$&$40$
\end{tabular}
\label{c1ek}
\end{table}

The CP asymmetries are also flavor dependent. 
They quantify the net number of each lepton flavor generated per $N_i$ decay, and are given by~\cite{Covi:1996wh}
\begin{align}
    \epsilon_{\alpha i}:=
    & 4\times 
    \frac{\Gamma(N_i \to \ell_{L_\alpha}+H_u) 
    -\Gamma(N_i\to {\overline{\ell}_{L_\alpha}}+H_u^\dag)}
    {\Gamma(N_i\to {\ell_L}+H_u) 
    + \Gamma(N_i\to {\overline{\ell}_L}+H_u^\dag)}\nonumber
    \\ = 
    &-\frac{1}{2\pi}\frac{1}{(y_{\nu_D}^\dag y_{\nu_D})_{ii}}
    \sum_{j\neq i}\Im\left[{(y_{\nu_D})}^\ast_{\alpha i} 
    (y_{\nu_D})_{\alpha j}(y_{\nu_D}^\dag y_{\nu_D})_{ij}\right] 
    f(M_j^2/M_i^2). 
\label{eq:flavoredCPasy}
\end{align}
In the hierarchical limit $M_1 \ll M_2, M_3$, one finds 
\begin{align}  
    \epsilon^{\ell_L}_{N_1}
    \simeq& 
    -\frac{3}{8\pi}\frac{1}{(y_{\nu_D}^\dag y_{\nu_D})_{11}} 
    \sum_{j}^3\Im\left[(y_{\nu_D}^\dag y_{\nu_D})^2_{1j}\right]
    \frac{M_1}{M_j}
    \propto 
    r_1.     
\label{eq:N1flavoredCPasy}
\end{align}
demonstrating that the $N_1$ contribution scales linearly with $r_1$. 
Table~\ref{c1ek} summarizes the flavored CP asymmetries $\epsilon_{\alpha i}$ and the decay parameters $K_{\alpha i}$ for $r_1=1$. 
The evaluation of $\epsilon_{\alpha i}$ proceeds in three steps.  
First, we compute ${(y_{\nu_D})}^\ast_{\alpha i}(y_{\nu_D})_{\alpha j}(y_{\nu_D}^\dag y_{\nu_D})_{ij}$ using the real Yukawa couplings determined by the underlying symmetry (Table~\ref{parameters}). 
Second, since all Yukawa couplings are assumed to be complex, we replace the imaginary part with the corresponding real part, $\Im\left[ {(y_{\nu_D})}^\ast_{\alpha i} 
(y_{\nu_D})_{\alpha j}(y_{\nu_D}^\dag y_{\nu_D})_{ij} \right] 
= \Re\left[ {(y_{\nu_D})}^\ast_{\alpha i} 
(y_{\nu_D})_{\alpha j}(y_{\nu_D}^\dag y_{\nu_D})_{ij} \right]$. 
Finally, substituting these expressions yields the explicit values of the flavored CP asymmetries.

The evolution of the lepton asymmetry is further constrained by equilibrated SM interactions. 
Sphalerons and Yukawa scatterings preserve the charges $\Delta_\alpha=B/3-L_\alpha$ ($\alpha=e,\mu,\tau$), with $B-L=\sum_\alpha \Delta_\alpha$. The Boltzmann equations governing $N_i$ and $\Delta_\alpha$ thus include redistribution terms encoding how lepton and Higgs asymmetries are mapped into $\Delta_\alpha$ through conversion coefficients $C^L_{\alpha\beta}$ and $C^{\tilde{H}_u}_\beta$: $Y_{\Delta {{\ell_L}_\alpha}}:=Y_{{\ell_L}_\alpha}-Y_{{\overline{\ell}_L}_\alpha} = -\sum_\beta C^L_{\alpha\beta}Y_{\Delta_\beta }$, $Y_{\Delta {\tilde{H}_u}}:=Y_{\tilde{H}_u}-Y_{\overline{\tilde{H}}_u}= -\sum_\beta C^{\tilde{H}_u}_{\beta}Y_{\Delta_\beta }$. 
The effect of Higgs asymmetry is so-called the spectator effect, which could affect the final amount of $B-L$ asymmetry by a factor of $\mathcal{O}(1)$~\cite{Buchmuller:2001sr}. 
For $10^8 \lesssim T \lesssim 10^{11}\,\mathrm{GeV} $, the relevant coefficients are~\cite{Fong:2010qh}
\begin{align}
  C^{\tilde{H}_u}=\frac{1}{358}\left(\begin{array}{ccc}
    37&52&52
\end{array}\right),\hspace{20pt}
  A^\ell=\frac{1}{2148}\left(\begin{array}{ccc}
  906&-120&-120\\
  -75&688&-28\\
  -75&-28&688
\end{array}\right)
\label{CHCL}.
\end{align}
At temperatures lower than $10^8$ GeV, additional down quark Yukawa interaction equilibrates for 
$\tan\beta\simeq7$–8 as favored in $SO(10)\times U(1)_A$. 
Thus, the conversion coefficients $C^L$ and $C^H$ become different from those in eq.(\ref{CHCL}). If $r_1\sim 1$, this effect may change our result.
In the present work, however, we 
neglect this effect because we are interested in larger $r_1$.

\section{
Numerical results
}
\label{Sec:NumericalRes}

In this section, we present a numerical analysis of the baryon asymmetry generated in the 
$SO(10) \times U(1)_A$ model. We begin by examining the dependence of the asymmetry on $M_1$ 
with a fixed neutrino Yukawa matrix. This serves as a first step toward identifying the 
favored parameter space in this framework. To further clarify the role of $M_1$ within the 
model structure, we then analyze three key aspects individually:
(a) the contribution of the second-lightest RH neutrino $N_2$, 
(b) the impact of flavor effects and spectator effect on leptogenesis, and
(c) alternative patterns of neutrino Yukawa matrices.
Finally, we compare the results with those obtained in the $E_6 \times U(1)_A$ scenario, 
highlighting the dependence of baryon asymmetry generation on the underlying GUT structure. 

\subsection{$M_1$ dependence}

\begin{figure}[t]
\centering
\includegraphics[width=0.6\linewidth]{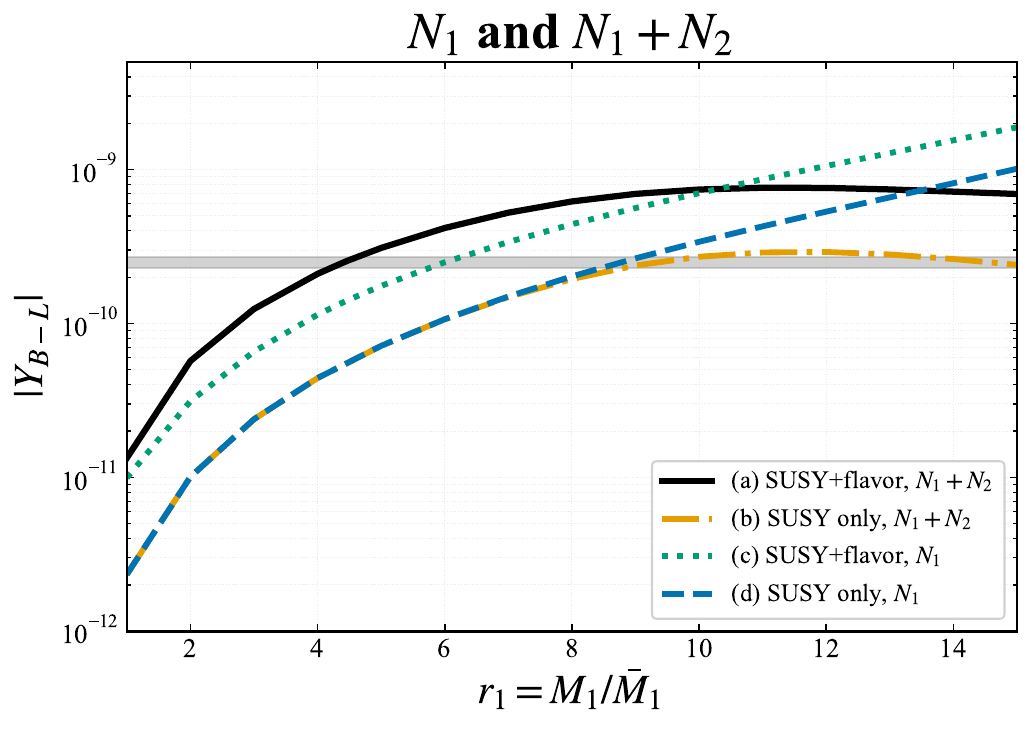}
\caption{
$r_1$ dependence of $Y_{B-L}$ for
(a) $N_1+N_2$ with the flavor effect, 
(b) $N_1+N_2$ without the flavor effect, 
(c) $N_1$ only with the flavor effect, 
(d) $N_1$ only without the flavor effect. 
}
\label{fig:r1Dep}
\end{figure}

Fig.~\ref{fig:r1Dep} shows the $r_1$ dependence of the $B-L$ asymmetry for
(a) $N_1+N_2$ with flavor effects, 
(b) $N_1+N_2$ without flavor effects, 
(c) $N_1$ only with flavor effects, and 
(d) $N_1$ only without flavor effects. 
The gray band indicates the required $B-L$ asymmetry that reproduces the observed baryon asymmetry. 
In the most realistic setup, namely $N_1+N_2$ with flavor effects, this corresponds to $r_1=4.4$. 
Such an enhancement is naturally achievable through $\mathcal{O}(1)$ variations in the model parameters.

We first consider the case without flavor effects. As $r_1$ increases, the generated $B-L$ asymmetry 
grows because the CP asymmetry scales as $\epsilon_1\propto r_1$, while the decay parameter decreases 
as $K_1\simeq 50/r_1$ and approaches unity. Since $K_1$ is typically large, the $N_2$ contribution is 
negligible for $M_1\ll M_2$, and becomes relevant only when $r_1\gtrsim 10$ where $M_1$ approaches $M_2$.

Turning to the flavored case, a qualitatively different picture emerges. The key point is that the 
flavored decay parameters, in particular $K_{e1}$ and $K_{\mu 1}$, can be of order unity or smaller, 
so the $N_2$ contribution cannot be neglected. As shown in Table~\ref{c1ek}, the flavored decay parameters 
at $r_1=1$ are already smaller than the unflavored $K_1$, and since $K_{\alpha 1}\propto 1/r_1$, these become even smaller when $r_1>1$.
Indeed, Fig.~\ref{fig:r1Dep} demonstrates that the $N_2$ effect remains 
significant even for $r_1=1$, largely because the $e$-flavor asymmetry retains sensitivity to its 
initial conditions because $K_{e1}=1.9<3$.

From this analysis, the favored value of the lightest RH-neutrino mass is determined as
$M_1 = 4.4 \times (2.6 \times 10^8\,\text{GeV}) \simeq 1.1 \times 10^9\,\text{GeV}$, 
consistent with the $SO(10)\times U(1)_A$ symmetry structure and the observed baryon asymmetry. 
This further implies a prediction for the lightest LH-neutrino mass of
$m_{\nu 1}\sim m_{\nu 1}(r_1=1)/r_1\sim \lambda^2 m_{\nu 3}/r_1\sim 5.5\times 10^{-4}$ eV. Here, we have used the relation between LH neutrino masses fixed by the symmetry as $m_{\nu_1}(r_1=1)\sim \lambda^2 m_{\nu_3}$\,\cite{Maekawa:2001uk, Maekawa:2002mx}, and the measured value, $m_{\nu3}\sim 0.05$ eV. 

\subsection{Evolution of $B-L$ asymmetry with $N_2$ (without flavor effect)} 
\label{Sec:N2}

\begin{figure}[h]
\centering
\includegraphics[width=0.6\linewidth]{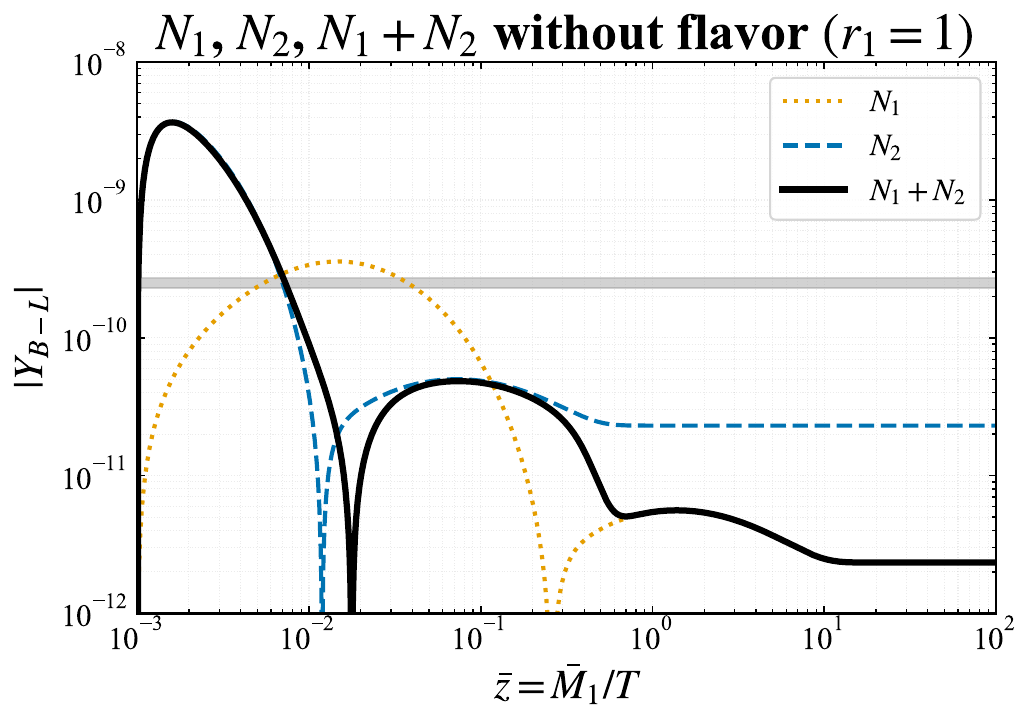}
\caption{$Y_{B-L}$ evolutions for $N_1+N_2$ (black solid), for $N_2$ only (blue dashed), and for $N_1$ only (yellow dotted). This calculation does not take the flavor effect into account. Gray band indicates the range of $B-L$ asymmetry required to reproduce the observed baryon asymmetry.}
\label{Fig:N1N2}
\end{figure}

In order to understand the results of the flavored leptogenesis, we demonstrate the evolution of $B-L$ asymmetry to study the effects of $N_2$ in this subsection.
Fig.~\ref{Fig:N1N2} shows the evolution of the $B-L$ asymmetry as a function of $\bar{z} := \overline{M}_1 / T$ for $r_1 = 1$ in three cases: [$N_1 + N_2$], [$N_1$ only], and [$N_2$ only]. 
Although the evolution differs clearly among these cases, the final $B-L$ asymmetry coincides in the [$N_1 + N_2$] and [$N_1$ only] cases. This behavior reflects the strong washout regime, where $K_1 \simeq 50 \gg 1$. 
At early times ($\bar{z} \lesssim 0.5$), the $B-L$ asymmetry is mainly washed out by $N_2$ interactions. For $\bar{z} \gtrsim 0.5$, however, washout induced by $N_1$ becomes dominant, driving the total $B-L$ asymmetry toward that of the [$N_1$ only] case.
To understand the result, we examine each stage of the evolution step by step. 

\

[$N_1$ only (yellow dotted)]
\\[-9mm]
\begin{enumerate}
\item At the early stage ($\bar{z} \simeq 0.001$), a $B-L$ asymmetry with the same sign as the CP asymmetry is generated during the thermal production of $N_1$.

\item Once $N_1$ approaches equilibrium ($0.01 \lesssim \bar{z} \lesssim 0.2$), the initially produced $B-L$ asymmetry is partially washed out by CP-conserving processes, such as $\ell_L \bar{N}_1 \to Q_L \bar{t}$, where $Q_L$ and $t$ denote the quark doublet and the singlet 
top quark, respectively.
Concurrently, CP-violating decays of $N_1$ generate a $B-L$ asymmetry with the opposite sign. 
As the evolution progresses, the positive contribution dominates, and from this point onward, the total $B-L$ asymmetry remains positive.

\item After $N_1$ decays ($\bar{z} \gtrsim 1$), the $B-L$ asymmetry experiences a slight reduction due to ongoing washout processes. Eventually ($\bar{z} \gtrsim 10$), the asymmetry reaches a constant value, as the rates of $B-L$-violating reactions drop below the Hubble expansion rate. 
This ``freeze-out" value is a conserved quantity in the SM.
\end{enumerate}

\

[$N_1 + N_2$ (black solid)]
\\[-9mm]
\begin{enumerate}
\item For $\bar{z} \lesssim 0.2$, the $B-L$ asymmetry is primarily determined by $N_2$ decays.
Since $\varepsilon_2$ is roughly an order of magnitude larger than $\varepsilon_1$ for similar 
decay parameters ($K_2 \simeq K_1$), the generated $B-L$ asymmetry exceeds that of the [$N_1$ only] 
scenario by about one order of magnitude.

\item For $\bar{z} \gtrsim 0.2$, the washout induced by $N_1$ becomes dominant ($K_1 \sim 50$), and 
the total $B-L$ asymmetry gradually converges toward the [$N_1$ only] result. 
It should be noted that if $K_1 \lesssim 3$, the asymmetry in the [$N_1 + N_2$] context would 
freeze out before fully reaching the [$N_1$ only] value, due to the weaker washout effect. 
\end{enumerate}

\subsection{Flavor effect on leptogenesis}

\begin{figure}[h]
\begin{tabular}{cc}
\begin{subfigure}[b]{0.49\columnwidth}
\centering
\includegraphics[width=\linewidth]{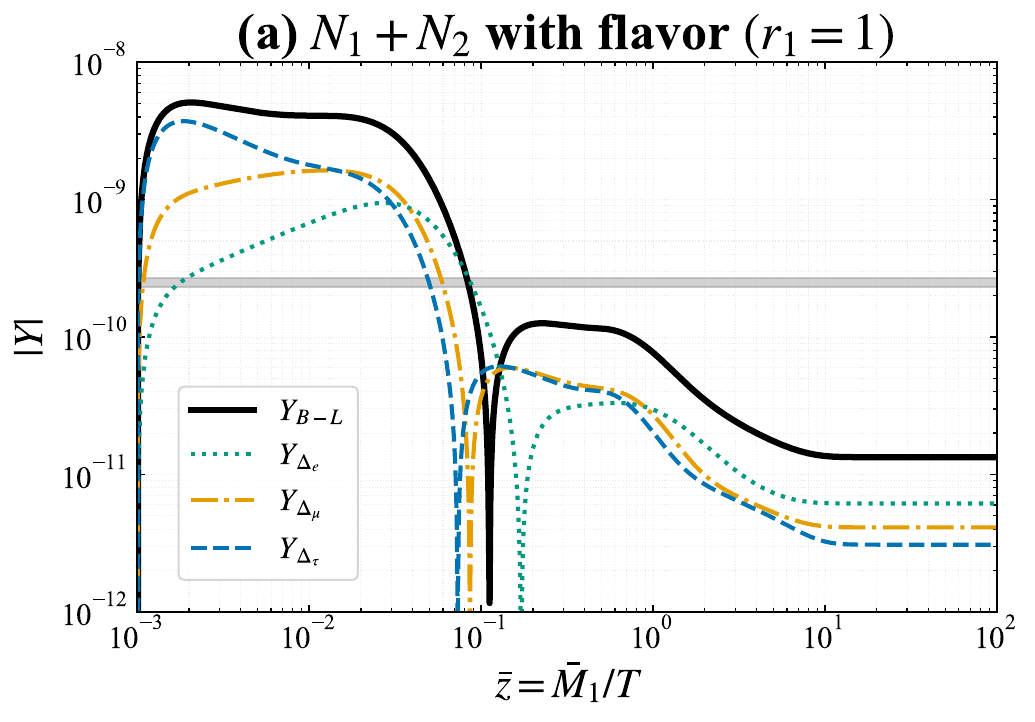}  
\end{subfigure}
&
\begin{subfigure}[b]{0.49\columnwidth}
\includegraphics[width=\linewidth]{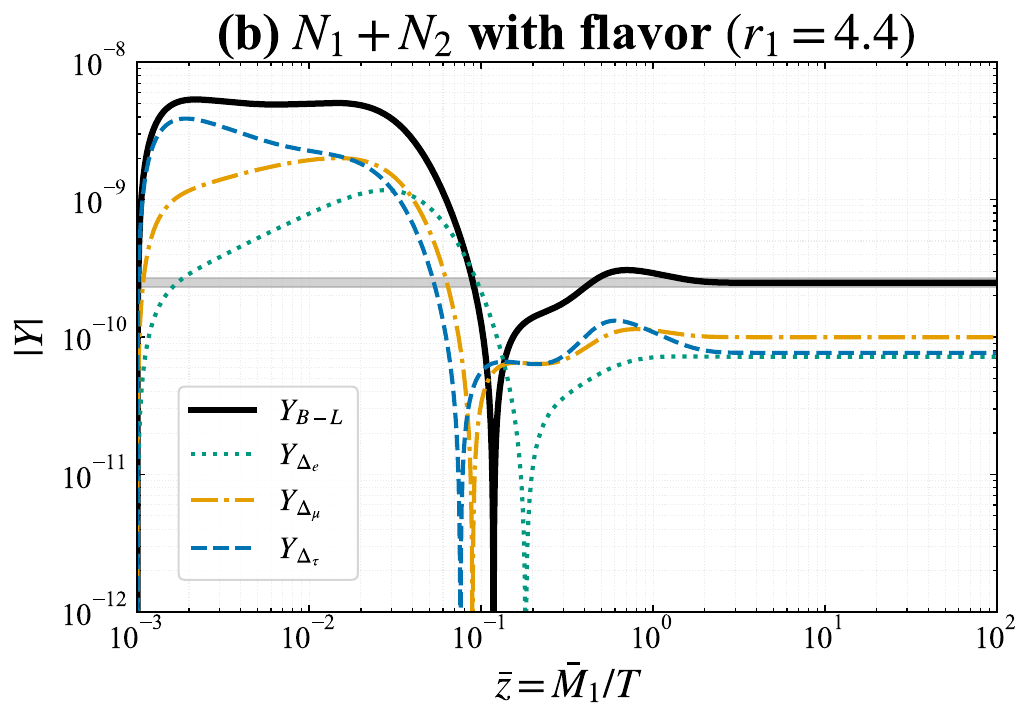}
\end{subfigure}
\end{tabular}
\caption{Evolutions of $Y_{B-L}$ and each flavor contribution for $r_1 = 1$ (a) 
and $4.4$ (b). Gray band indicates the range of $B-L$ asymmetry required to 
reproduce the observed baryon asymmetry.}
\label{Fig:flavor-r1-r4.4}
\end{figure}

In this section, we numerically illustrate the impact of flavor effects and spectator effect within our model parameters.
Comparing the black solid line in the left panel of Fig.~\ref{Fig:flavor-r1-r4.4} with the black 
solid line in Fig.~\ref{Fig:N1N2}, we observe that the final $B-L$ asymmetry is enhanced by roughly 
an order of magnitude.
This increase is mainly due to the reduced washout arising from the smaller flavored decay parameters, 
$1 < K_e < K_\mu < K_\tau < K_1$. Consequently, the individual flavored asymmetries satisfy 
$Y_{\Delta_e} > Y_{\Delta_\mu} > Y_{\Delta_\tau} > Y_{B-L}^{\rm without\ flavor}$, even though the 
corresponding CP asymmetries follow $\epsilon_e < \epsilon_\mu < \epsilon_\tau < \epsilon_1$, 
as summarized in Table~\ref{c1ek}.

The right panel in Fig.~\ref{Fig:flavor-r1-r4.4} shows the same quantities as the left panel but 
for $r_1 = 4.4$. Compared to $r_1 = 1$, the flavored decay parameters $K_{\alpha 1}$
are reduced as 
$K_{e1} < 1 \sim K_{\mu 1} < K_{\tau 1} \sim 8$, and the CP asymmetries $\epsilon_{\alpha 1}$ are larger with the relations
$\epsilon_{e1} < \epsilon_{\mu 1} < \epsilon_{\tau 1}$.
As a result, the muon and tau asymmetries are enhanced, while the electron asymmetry is mainly 
determined by $N_2$ decays, where $\epsilon_{e2} > \epsilon_{e1}$ and $K_{e2} \sim 2$, since the 
washout from $N_1$ is weak for this flavor. These effects together allow the total $B-L$ asymmetry 
to reach a level consistent with the observed baryon number.

Fig.~\ref{Fig:flavor-r1-r4.4} also illustrates the $\overline{z} := \overline{M}_1/T$ evolution 
of the flavored asymmetries for $N_1+N_2$ with flavor effects. For $r_1 = 1$, the strong washout 
regime leads to a marked suppression of the muon and tau asymmetries, yielding a total $B-L$ below 
the observed value. 
In contrast, $r_1 = 4.4$ enhances $\epsilon_{\alpha 1}$ and reduces $K_\mu$ and $K_\tau$ toward unity, 
resulting in the correct baryon asymmetry. Values of $r_1$ above 4.4 further increase the tau 
contribution, causing the total $B-L$ to exceed the required value. Thus, $M_1$ is a crucial parameter 
controlling both the flavor effects and the magnitude of the flavored CP asymmetries, as reflected in 
eq.(\ref{Eq:K_alphai}) and eq.(\ref{eq:N1flavoredCPasy}).

\subsection{Other patterns of neutrino Yukawa matrices}

We so far have  
computed the flavored CP asymmetries given in Eq. (\ref{eq:flavoredCPasy}) under 
the following assumptions. Using the real Yukawa couplings 
fixed by symmetry (Table~\ref{parameters}) with all $\mathcal{O}(1)$ coefficients set to unity, 
we have evaluated 
${(y_{\nu_D})}^\ast_{\alpha i} (y_{\nu_D})_{\alpha j}(y_{\nu_D}^\dag y_{\nu_D})_{ij}$. 
Since the couplings are generically complex, we have taken
$\Im\left[{(y_{\nu_D})}^\ast_{\alpha i}  
(y_{\nu_D})_{\alpha j}(y_{\nu_D}^\dag y_{\nu_D})_{ij}\right]=\Re\left[{(y_{\nu_D})}^\ast_{\alpha i} 
(y_{\nu_D})_{\alpha j}(y_{\nu_D}^\dag y_{\nu_D})_{ij}\right]$.
This yields maximal CP asymmetries, because contributions from virtual RH neutrinos 
add constructively. In general, cancellations among complex coefficients can reduce the asymmetries. 
To explore this, we also consider scenarios where some Yukawa entries have $-1$ as their 
$\mathcal{O}(1)$ coefficient.

Flipping signs in the neutrino Yukawa matrix changes the flavored CP asymmetries $\epsilon_{i\alpha}$ 
in magnitude and sign, while the decay parameters remain unchanged.
Table~\ref{tableepsilonsign} 
summarizes all patterns, which can be classified into 15 patterns (see Appendix B). 
\begin{table}[h]
  \centering\caption{Corrected version of table 4 in the published paper.
The correlation between the sign pattern of each component of the Yukawa matrix \( y_{\nu_D} \) and resulting sign of the flavored CP asymmetry 
\( \epsilon_{\alpha i} \), and the corresponding maximum value of the $B-L$ asymmetry $Y_{B-L}$. 
From left to right, the columns show
(1) the components whose signs are flipped, 
(2) the signs of \( \epsilon_{e1}, \epsilon_{\mu1}, \epsilon_{\tau1}, 
\epsilon_{e2}, \epsilon_{\mu2}, \epsilon_{\tau2} \) for \( r_1 = 1 \), 
(3) the signs of \( \epsilon_{e1}, \epsilon_{\mu1}, \epsilon_{\tau1}, 
\epsilon_{e2}, \epsilon_{\mu2}, \epsilon_{\tau2} \) for \( r_1 = 15 \), 
(4) the maximum value of \( R(r_1):= Y_{B-L}(r_1) / Y^{\text{obs}}_{B-L} \), 
and (5) $r_1$ accounting for $Y^{\text{obs}}_{B-L}$. 
The values of \(R_{\max}\) are rounded to two significant figures, while the existence of \(r_1^{\rm obs}\) is determined using the unrounded numerical values.
}
\begin{tabular}{l||l|l|ll|l}
  Flipped component(s)&$\epsilon_{1\alpha},\epsilon_{2\alpha}(r_1=1)$&$\epsilon_{1\alpha},\epsilon_{2\alpha}(r_1=15)$&$R^{\text{max}}$&$(r_1^{\text{max}})$&$r_1^{obs}$\\
  \hline
  \hline
  $0$.(all positive)&$---$,$---$&&$3.0$&$(11)$&$4.4$\\
  $1$.[(${y_{\nu_D}})_{e1}],\,$...&$+--$, $---$&&$2.0$ &$(11)$&$6\sim 7$\\
  $2$.$[({y_{\nu_D}})_{\mu 1}],\,$...&$-+-$,$---$&&$0.93$  &$(12)$&\\
  $3$.$[({y_{\nu_D}})_{\tau 1}],\,$...&$++-$,$---$&&$0.30$  &$(12)$&\\
  $4$.$[({y_{\nu_D}})_{e2}],\,$...&$---$, $+--$&$+--$, $+--$&$2.2$ &$(11)$&$5\sim6$\\
  $5$.$[({y_{\nu_D}})_{\mu2}],\,$...&$---$, $-+-$&$-+-$, $-+-$&$1.7$ &$(11)$&$6\sim7$\\
  $6$.$[({y_{\nu_D}})_{\tau2}],\,$...&$---$, $++-$&&$1.2$ &$(10)$&$8\sim9$\\
  $7$.$[({y_{\nu_D}})_{e3}],\,$...&$---$, $+--$&&$2.3$ &$(11)$&$5\sim6$\\
  $8$.$[({y_{\nu_D}})_{\mu3}],\,$...&$---$, $-+-$&&$1.9$  &$(12)$&$6\sim7$\\
  $9$.$[({y_{\nu_D}})_{\tau3}],\,$...&$---$, $++-$&&$1.5$ &$(12)$&$7\sim8$\\
  $10$.$[({y_{\nu_D}})_{e1},({y_{\nu_D}})_{\mu2}],\,$...&$+--$, $-+-$&$++-$, $-+-$&$0.86$ &$(10)$&\\
  $11$.$[({y_{\nu_D}})_{e1},({y_{\nu_D}})_{\tau2}],\,$...&$+--$, $++-$&$-+-$, $++-$&$1.0$ &$(11)$&$9\sim10$\\
  $12$.$[({y_{\nu_D}})_{e1},({y_{\nu_D}})_{\mu3}],\,$...&$+--$, $-+-$& &$1.0$ &$(11)$&$10\sim11$\\
  $13$.$[({y_{\nu_D}})_{e1},({y_{\nu_D}})_{\tau3}],\,$...&$+--$, $++-$&&$1.1$ &$(11)$&$8\sim9$\\
  $14$.$[({y_{\nu_D}})_{\mu1},({y_{\nu_D}})_{e2}],\,$...&$-+-$, $+--$&$++-$, $+--$&$0.30$ &$(11)$&\\
  $15$.$[({y_{\nu_D}})_{\mu1},({y_{\nu_D}})_{e3}],\,$...&$-+-$, $+--$&&$0.38$ &$(12)$&\\
  \hline
 \end{tabular}
  
\label{tableepsilonsign}
\end{table}
The leftmost column indicates which Yukawa components are 
flipped. Columns two and three show the CP asymmetry signs for each flavor, including $r_1=15$ when 
they differ from $r_1=1$. Column four gives
$R^{\text{max}} = Y_{B-L}(r_1^{\text{max}})/Y_{B-L}^{\text{obs.}}$, 
where the $B-L$ asymmetry is maximal, and column five lists the corresponding $r_1$ required to 
reproduce the observed baryon asymmetry.

\begin{figure}[h!]
\centering
\begin{tabular}{cc}
\begin{subfigure}[b]{0.49\columnwidth}
\centering
\includegraphics[width=\linewidth]{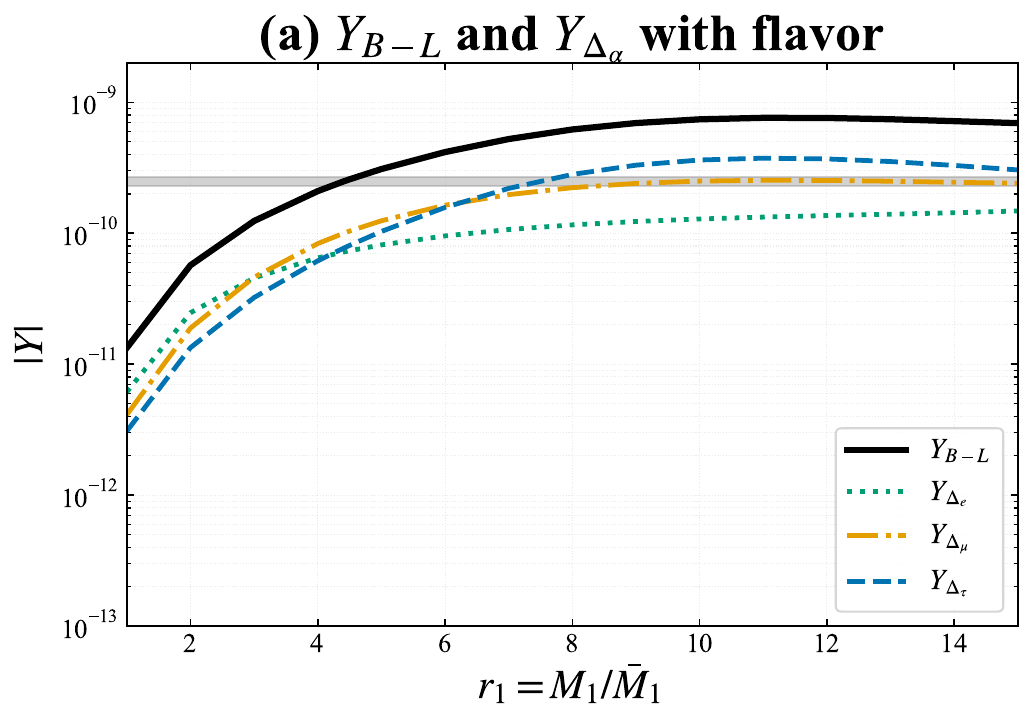}  
\end{subfigure}
&
\begin{subfigure}[b]{0.49\columnwidth}
\includegraphics[width=\linewidth]{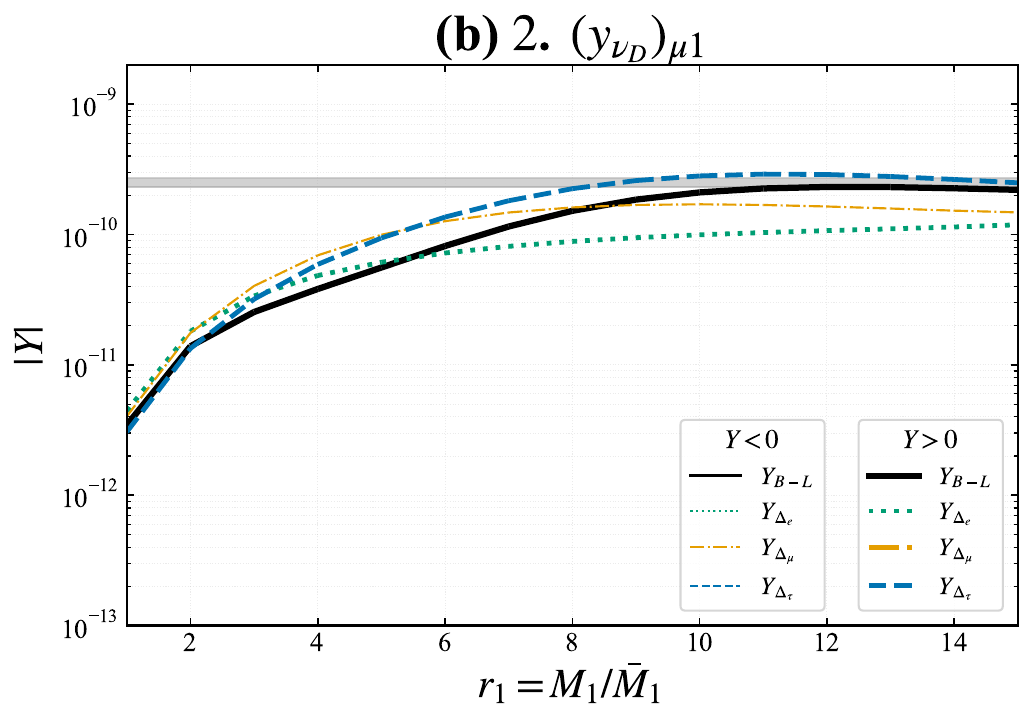}
\end{subfigure}\\
\begin{subfigure}[b]{0.49\columnwidth}
\centering
\includegraphics[width=\linewidth]{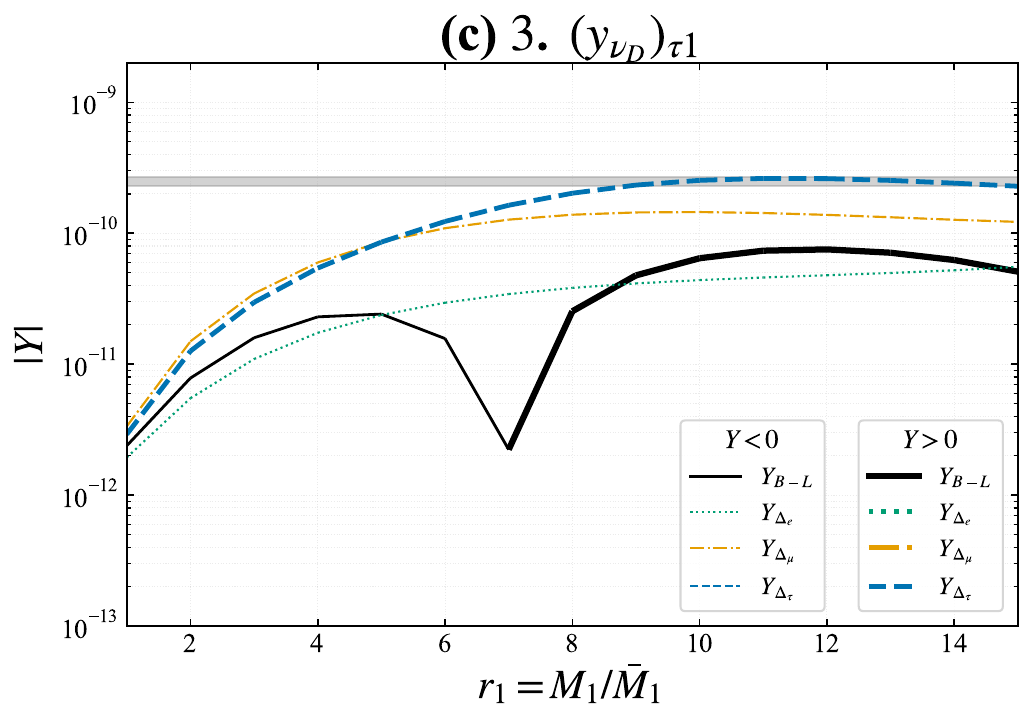}  
\end{subfigure}
&
\begin{subfigure}[b]{0.49\columnwidth}
\includegraphics[width=\linewidth]{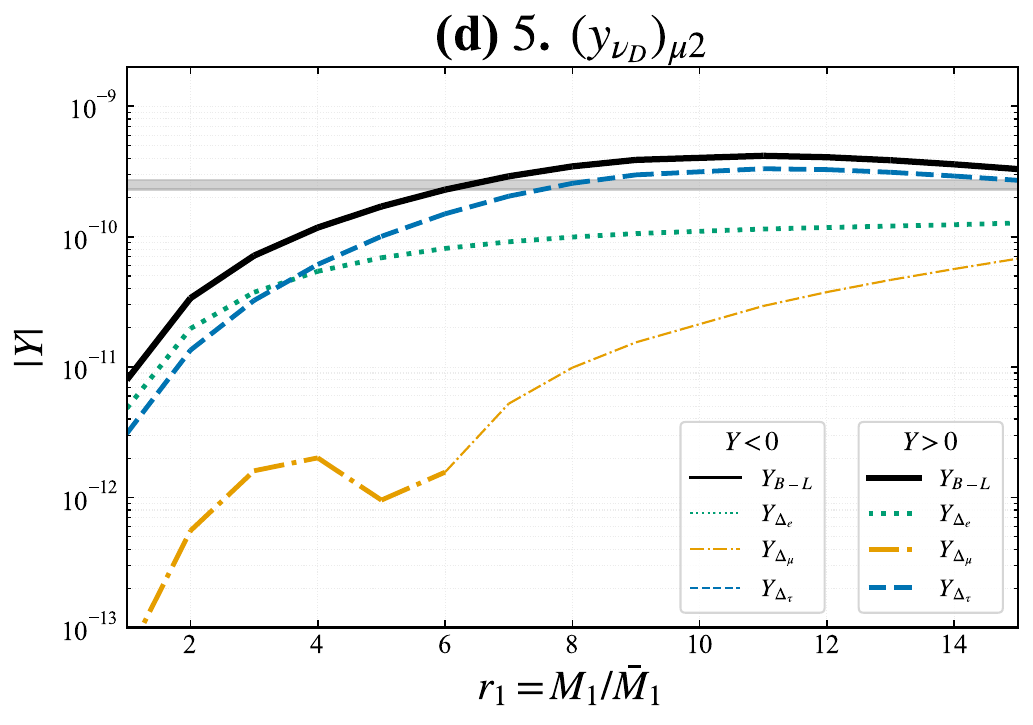}
\end{subfigure}
\end{tabular}
\caption{
Corrected version of figure 4 in the published paper.
The $r_1$ dependence of the total and flavored asymmetries is shown in four cases:
(a) pattern 0,
(b) pattern 2,
(c) pattern 3,
and (d) pattern 5.
The absolute values of the asymmetries are plotted.
In contrast to the published figure, the sign of each flavored asymmetry
$Y_{\Delta_\alpha}$ is now encoded by the line width:
thin lines denote $Y_{\Delta_\alpha}<0$, while thick lines denote
$Y_{\Delta_\alpha}>0$.
The line styles and colors distinguish $Y_{B-L}$,
$Y_{\Delta_e}$, $Y_{\Delta_\mu}$, and $Y_{\Delta_\tau}$.
}
\label{yukawa_graph}
\end{figure}

All patterns with negative $\epsilon_{1\alpha}$ can achieve the observed baryon asymmetry. 
In these cases, all flavored contributions $\Delta_\alpha$ add constructively (Fig.~\ref{yukawa_graph}(a) and (d)), although two of them (patterns 4 and 5) change the sign of $\epsilon_{1\alpha}$ between $r_1=1$ and $15$. 
In pattern 5, $\Delta_\mu$ flips sign near $r_1 \sim 6$ due to the sign-change of $\epsilon_{1\mu}$. 
Roughly half of the patterns with mixed-sign $\epsilon_{1\alpha}$ cannot reach the observed asymmetry because of cancellations among flavored contributions (Fig.~\ref{yukawa_graph}(b) and (c)).
In pattern 3, for example, the total $B-L$ asymmetry changes sign around $r_1 \sim 7$. 
Overall, ten of fifteen patterns can achieve the observed asymmetry with certain $r_1<15$; 
the other five patterns yield $B-L$ of similar order but below the observed value.

Our calculations may underestimate $B-L$, since $\pm 1$ coefficients enhance cancellations. 
For instance, in pattern 5 with $[(y_{\nu_D})_{\mu 2}]$ flipped, 
\begin{align}
    \epsilon_{\mu1}\propto 
    &-\Im \left[ 
    \left(y_{\nu_D}\right)_{\mu1}^\ast   \left(y_{\nu_D}\right)_{\mu2} 
    \left\{\left(y_{\nu_D}\right)^\ast_{e1}   \left(y_{\nu_D}\right)_{e2}  
    -\left(y_{\nu_D}\right)^\ast_{\mu1}   \left(y_{\nu_D}\right)_{\mu2} 
    +\left(y_{\nu_D}\right)^\ast_{\tau1}  \left(y_{\nu_D}\right)_{\tau2}\right\} 
    \right] 
    f\left(\frac{M^2_2}{M^2_1}\right)
    \nonumber
    \\
    &+\Im\left[ 
    \left(y_{\nu_D}\right)_{\mu1}^\ast   \left(y_{\nu_D}\right)_{\mu3}  
    \left\{\left(y_{\nu_D}\right)^\ast_{e1}    \left(y_{\nu_D}\right)_{e3}  
    +\left(y_{\nu_D}\right)^\ast_{\mu1}   \left(y_{\nu_D}\right)_{\mu3} 
    +\left(y_{\nu_D}\right)^\ast_{\tau1} \left(y_{\nu_D}\right)_{\tau3}\right\}
    \right] 
    f\left(\frac{M^2_3}{M^2_1}\right)
    \\
    = 
    &-\Im\left[ 
    \left(y_{\nu_D}\right)_{\mu1}^\ast  
    \left(y_{\nu_D}\right)_{\mu2}  
    (y_{\nu_D}^\dag y_{\nu_D})_{12}
    \right]
    f\left(\frac{M^2_2}{M^2_1}\right)
    +\Im\left[ 
    \left(y_{\nu_D}\right)_{\mu1}^\ast 
    \left(y_{\nu_D}\right)_{\mu3} 
    (y_{\nu_D}^\dag y_{\nu_D})_{13}
    \right]
    f\left(\frac{M^2_3}{M^2_1}\right)
    \nonumber\\
    &+2\Im\left[ 
    \left\{\left(y_{\nu_D}\right)^\ast_{\mu1} 
    \left(y_{\nu_D}\right)_{\mu2}\right\}^2 
    \right]f\left(\frac{M^2_2}{M^2_1}\right), 
\label{epsflip}
\end{align} 
where all $\left(y_{\nu_D}\right)_{\alpha i}$ are given in Table \ref{parameters} with positive signs.
When $M_2, M_3 \gg M_1$, the functions $f\left(\frac{M_i^2}{M_1^2}\right) (i=2,3)$ become approximately $M_1/M_i$. 
Thus, in eq.(\ref{epsflip}) the first two terms ($\sim \lambda^{22}$) completely cancel, leaving the positive third term ($\sim \lambda^{24}$) to dominate.
Since this cancellation arises from the $\mathcal{O}(1)$ coefficients taking values of ±1, the CP asymmetry may be underestimated owing to a stronger cancellation compared with that for general complex Yukawa couplings.
Moreover, the above consideration can explain the sign-flips of several $\epsilon_{\alpha i}$ between $r_1=1$ and $15$. 
This is because the accuracy of the above approximation $M_1\ll M_{2}$ decreases with larger $r_1$. 
As a result, the first term in eq.(\ref{epsflip}) dominates, and the overall sign of $\epsilon_{\mu 1}$ can flip from positive to negative for larger $r_1$, explaining the sign flip of $Y_{\Delta_\mu}$ around $r_1 \sim 6$ in Fig.\ref{yukawa_graph}(d). 
Similar reasoning applies to other $\epsilon_{\alpha i}$ 
flips between $r_1=1$ and $r_1=15$.

Fig.~\ref{yukawa_graph} also shows representative $B-L$ evolutions. Across patterns, $B-L$ increases with 
$r_1$ and saturates around $r_1 \sim 10$. At large $r_1$, the tau flavor often dominates due to larger 
$\epsilon_\tau$ and $K_\tau$ approaching unity. Saturation is likely caused by quasi-degeneracy of 
$N_1$ and $N_2$, consistent with Ref.~\cite{Antusch:2011nz}, which notes significant effects for $M_2 \lesssim 3M_1$, that corresponds to $r_1\gtrsim 7$.


\section{Comparison with $\mathbf{E_6 \times U(1)_A}$ model} 
\label{Sec:Discussion}

\begin{figure}[h]
\centering
\includegraphics[width=0.6\linewidth]{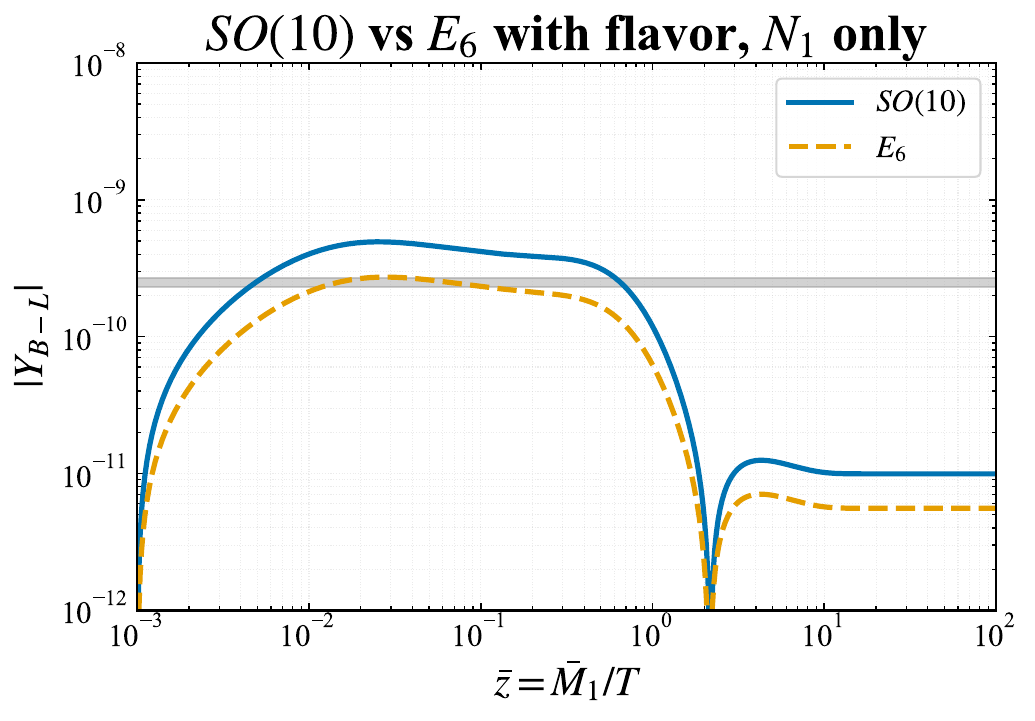}
\caption{Evolution of $Y_{B-L}$ in the $SO(10) \times U(1)_A$ and $E_6 \times U(1)_A$ scenarios. 
We set $r_1=1$ in the $N_1$ framework including flavor effects. 
The grey band denotes the required $B-L$ asymmetry to account for the observed baryon asymmetry.} 
\label{Fig:E6SO10}
\end{figure}

GUT models based on the $E_6$ gauge group are also attractive, with the $E_6 \times U(1)_A$ scenario being particularly successful from a phenomenological perspective~\cite{Bando:2001bj, Maekawa:2002bk}. 
Before summarizing this work, we compare the dependence of baryon asymmetry generation on the choice of GUT scenario, namely $SO(10) \times U(1)_A$ versus $E_6 \times U(1)_A$ ~\cite{Ishihara:2015uua}.

\begin{table}[h]
\centering
\caption{
Masses of the lightest and heaviest RH neutrinos, and $ij$ components of $y_{\nu_D}^\dag y_{\nu_D}$. 
For simplicity, all $\mathcal{O}(1)$ coefficients are set to $1$.
}
\begin{tabular}{c||c|c|c|c|c}
    & lightest $N_i$ mass  & heaviest $N_i$ mass& $(y_{\nu_D}^\dag y_{\nu_D})_{11}$ & $(y_{\nu_D}^\dag y_{\nu_D})_{13}$ & $(y_{\nu_D}^\dag y_{\nu_D})_{16}$ \\
    \hline\hline
    $SO(10)$ & $M_1\sim\lambda^{12}\Lambda_G$ & $M_3\sim\lambda^6\Lambda_G$
    & $\lambda^{10}$ & $\lambda^{7}$ & \\
    $E_6$ & $M_1\sim\lambda^{13}\Lambda_G$ & $M_6\sim\lambda^6\Lambda_G$ &
    $\lambda^{11}$ & $\lambda^{10}$ & $\lambda^{7.5}$ \\
\end{tabular}
\label{E5SO10m1}
\end{table}

The evolution of the $B-L$ asymmetry for $r_1=1$ in these scenarios is shown in Fig.~\ref{Fig:E6SO10}, where flavor effects are included. 
For simplicity, all $\mathcal{O}(1)$ coefficients are set to $1$. 
Within this setup, the $SO(10) \times U(1)_A$ scenario produces approximately twice the $B-L$ asymmetry of the $E_6 \times U(1)_A$. 
This conclusion holds whether or not flavor effects are included. 

The $E_6\times U(1)_A$ model has six RH neutrinos, and the mass of the lightest RH neutrino is smaller than that in the $SO(10)\times U(1)_A$ case, as shown in Table~\ref{E5SO10m1}. 
However, the flavored decay parameters in the $E_6\times U(1)_A$ model are of the same order in $\lambda$ as those in the $SO(10)\times U(1)_A$ model (see Table~\ref{c1ek}), because the Dirac neutrino Yukawa couplings of the lightest RH neutrino are smaller. 

The key ingredient in understanding the factor-of-two difference between the scenarios is the CP asymmetry, which follows from the symmetries in each models as
\begin{equation}
\begin{split}
    \left|\epsilon^{SO(10)}_1\right| 
    & \sim 
    \frac{3}{8\pi}\times 2 \times 
    \frac{\Im\!\left[(y_{\nu_D} ^\dag y_{\nu_D})^2_{13}\right]}{(y_{\nu_D} ^\dag y_{\nu_D})_{11}} 
    \frac{M_1}{M_3}
    \sim \frac{3}{4\pi}\lambda^{10},
    \\
    \left|\epsilon^{E_6}_1\right| 
    & \sim 
    \frac{3}{8\pi}\times 5 \times 
    \frac{\Im\!\left[(y_{\nu_D} ^\dag y_{\nu_D})^2_{16}\right]}
    {(y_{\nu_D} ^\dag y_{\nu_D})_{11}}\frac{M_1}{M_6} 
    \sim \frac{15}{8\pi}\lambda^{11}\;\sim 0.55\, \epsilon^{SO(10)}_1, 
\label{Eq:SO10E6-CPasy}
\end{split}     
\end{equation}
where the relevant RH-neutrino masses and Yukawa combinations are summarized in Table~\ref{E5SO10m1}, and detailed parameter values in $E_6\times U(1)_A$ GUT are given in Ref.~\cite{Ishihara:2015uua}. 

Note that the $U(1)_A$ symmetry ensures the universality of $\lambda$ dependence in the combinations of couplings and masses involving virtual RH neutrinos, i.e., 
\[
\frac{\Im[(y_{\nu_D} ^\dag y_{\nu_D})^2_{1i}]}{(y_{\nu_D} ^\dag y_{\nu_D})_{11}} \cdot 
\frac{M_1}{M_i} \propto \lambda^{10} \ \text{in } SO(10),\quad 
\propto \lambda^{11} \ \text{in } E_6,
\]
for any generation $i$. 
Accordingly, we first estimate the CP asymmetry considering only the heaviest RH neutrino, and then multiply by a factor counting the number of generations of virtual RH neutrinos: 2 in $\epsilon^{SO(10)}_1$ and 5 in $\epsilon^{E_6}_1$, respectively
\footnote{In the literature on $E_6$ leptogenesis, a counting factor $\sqrt{4}=2$ is sometimes used instead of 5. 
Here the factor 4 counts the internal-loop RH neutrinos $N_3, N_4, N_5$, 
and $N_6$, and the square root reflects the randomness of signs in the neutrino Yukawa couplings~\cite{Ishihara:2015uua}.}. 

Even when flavor effects are included, this reasoning holds, since each flavor component of the neutrino Yukawa matrices scales with $\lambda$ in the same way in both scenarios. 
Consequently, we indicate that the $SO(10) \times U(1)_A$ scenario tends to yield a larger baryon asymmetry than the $E_6 \times U(1)_A$ case with fixed $r_1$, owing to the milder suppression of the CP asymmetry with respect to $\lambda$.

\section{Conclusion}
\label{Sec:Conclusion}

We have studied leptogenesis in the natural $SO(10)$ GUT ($SO(10)\times U(1)_A$ GUT), in which the doublet–triplet splitting problem is solved and the observed quark and lepton masses and mixings can be reproduced under the natural assumption that all interactions, including higher-dimensional ones, appear with $\mathcal{O}(1)$ coefficients. 
In the natural GUT, this assumption implies that everything is determined by the symmetry except for the $\mathcal{O}(1)$ coefficients. 

To obtain the predicted value of the baryon asymmetry in the natural $SO(10)$ GUT, we have to fix the $\mathcal{O}(1)$ coefficients. In this paper, to account for the observed baryon asymmetry, we have used the $\mathcal{O}(1)$ coefficient for $M_1$  while all $\mathcal{O}(1)$ coefficients of the Dirac neutrino Yukawa couplings are taken to be unity. 
We showed that the observed baryon asymmetry can be obtained when the coefficient $r_1$ is taken to be 4.4, corresponding to $M_1\sim 1.1\times 10^9\,\mathrm{GeV}$. 
This result suggests that the mass of the lightest LH neutrino, $m_{\nu_1}$, is reduced to $1/4.4$ of its symmetry-determined value, yielding $m_{\nu_1}\sim 5.5\times 10^{-4}\,\mathrm{eV}$.

We included flavor effects and the contribution of the second-lightest RH neutrino $N_2$ in our calculation. 
As a result, we reconfirmed the importance of $N_2$ in leptogenesis with flavor effects. 
In particular, the $N_2$ contribution becomes significant when some of the flavored decay parameters of the lightest RH neutrino are smaller than 3, or when $3M_1 > M_2$, which is consistent with the results of Ref.~\cite{Antusch:2011nz}.

In our calculation, we estimated the magnitudes of the flavored CP asymmetries using real Yukawa couplings determined by the symmetry, with $\mathcal{O}(1)$ coefficients set to unity, rather than directly evaluating the imaginary part of the expressions with complex Yukawa couplings. 
In this case, the flavored CP asymmetries are relatively large due to multiple additive contributions.  
However, with complex $\mathcal{O}(1)$ coefficients, cancellations may occur in the flavored CP asymmetries. 
We examined this cancellation effect by flipping the signs of the Yukawa couplings and showed that in ten of the 15 possible sign patterns, the observed baryon asymmetry can be realized. 
Even in the remaining patterns, the predicted baryon asymmetry, which does not reach to the observed value, becomes the same order of the observed baryon asymmetry by taking a larger $r_1<15$.

\section*{Acknowledgments}
This work is supported in part by the Grant-in-Aid for Scientific Research 
  from the Ministry of Education, Culture, Sports, 
 Science and Technology in Japan  No.~19K03823, 25K07321 (N.M.), 22K03638, 22K03602, 
 and 20H05852 (M.Y.).
 This work was financially supported by JST SPRING, Grant Number JPMJSP2125. The author (K.S.) would like to take this opportunity to thank the “THERS Make New Standards Program for the Next Generation Researchers.” 
\newpage
\clearpage
\newpage
\appendix
\section{Boltzmann Equations}
In this appendix, we write down the Boltzmann equations used in our calculation of the $B-L$ asymmetry, which include flavor effects and spectator effects. These equations are basically obtained in \cite{Fong:2010qh}.

The evolution of the $B-L$ asymmetry is governed by the Boltzmann equations for the right-handed neutrinos $N_1, N_2$ and the flavored $B-L$ asymmetries $\Delta_\alpha$. The scattering terms include the (inverse-) decays and scattering processes of these species. 
Diagrams containing multiple Dirac neutrino Yukawa couplings are neglected, since these couplings are at most of order $\mathcal{O}(\lambda^2)\sim 0.048$ in the natural $SO(10)$ GUT. 

Introducing the notation $\dot{Y} := sHz\, dY/dz$, the Boltzmann equations can be written as

\begin{align}
  -\dot{Y}_{N_i}=&\left(\frac{Y_{N_i}}{Y^{eq}_{N_i}}-1\right)\left( \gamma_{N_i}+4\gamma^{eq(0)}_{t_i}+4\gamma^{eq(1)}_{t_i}+4\gamma^{eq(2)}_{t_i}+2\gamma^{eq(3)}_{t_i}+4\gamma^{eq(4)}_{t_i} \right),\\
  -\dot{Y}_{+i}=&\left(\frac{Y_{+_i}}{Y^{eq}_{\widetilde{N}_i}}-1\right)\left( \gamma_{\widetilde{N}_i}+3\gamma_{22_i}+2\gamma^{eq(5)}_{t_i}+2\gamma^{eq(6)}_{t_i}+2\gamma^{eq(7)}_{t_i}+\gamma^{eq(8)}_{t_i}+2\gamma^{eq(9)}_{t_i} \right),\\  
  -\dot{Y}_{-i}=&\frac{Y_{-i}}{Y^{eq}_{\widetilde{N}^c_i}}\left( \gamma_{\widetilde{N}_i}+3\gamma_{22_i}+2\gamma^{eq(5)}_{t_i}+2\gamma^{eq(6)}_{t_i}+2\gamma^{eq(7)}_{t_i}+\gamma^{eq(8)}_{t_i}+2\gamma^{eq(9)}_{t_i} \right)\nonumber\\
  &-\frac{Y_{\Delta \ell}}{Y^{eq}_{\ell}}\left( \frac{Y_{+i}}{Y^{eq}_{\widetilde{N}_i}}\gamma^{eq(5)}_{t_i}+2\gamma^{eq(6)}_{t_i}+2\gamma^{eq(7)}_{t_i}\right)\nonumber\\
  &-\frac{ Y_{\Delta \widetilde{\ell}}}{Y^{eq}_{\widetilde{\ell}}}\left[\left( 2+\frac{1}{2}\frac{Y_{+i}}{Y^{eq}_{\widetilde{N}_i}} \right)\gamma_{22_i} -\frac{1}{2}\frac{Y_{+i}}{Y^{eq}_{\widetilde{N}_i}}\gamma^{eq(8)}_{t_i}-2\gamma^{eq(9)}_{t_i}\right]\nonumber\\
  &-2\frac{Y_{\Delta \tilde{H}_u}}{Y^{eq}_{\widetilde{H}_u}}\left[ \left(2+\frac{1}{2}\frac{Y_{+i}}{Y^{eq}_{\widetilde{N}_i}}\right)\gamma_{22_i}+\gamma^{eq(5)}_{t_i}+\frac{1}{2}\frac{Y_{+i}}{Y^{eq}_{\widetilde{N}_i}}\gamma^{eq(6)}_{t_i}+\gamma^{eq(7)}_{t_i} \right]\nonumber\\
  &-\frac{Y_{\Delta H_u}}{Y^{eq}_{H}}\left[ \left(2+\frac{1}{2}\frac{Y_{+i}}{Y^{eq}_{\widetilde{N}_i}}\right)\gamma_{22_i}+2\gamma^{eq(5)}_{t_i}+\left(1+\frac{1}{2}\frac{Y_{+i}}{Y^{eq}_{\widetilde{N}_i}}\right)\left( \gamma^{eq(6)}_{t_i}+\gamma^{eq(7)}_{t_i} \right)-\gamma^{eq(8)}_{t_i}-2\gamma^{eq(9)}_{t_i} \right]  ,
\end{align}

\begin{align}
-\dot{Y}_{\Delta \alpha}=\sum_i &\left[ \epsilon_{\alpha i}\left( \frac{Y_{N_i}}{Y^{eq}_{N_i}}-1 \right)\left( \gamma_{N_i}+4\gamma^{eq(0)}_{t_i}+4\gamma^{eq(1)}_{t_i}+4\gamma^{eq(2)}_{t_i}+2\gamma^{eq(3)}_{t_i}+4\gamma^{eq(4)}_{t_i} \right)\right.\nonumber\\
  &+\epsilon_{\alpha i}\left(\frac{Y_{+_i}}{Y^{eq}_{\widetilde{N^c}_i}}-1\right)\left( \gamma_{\widetilde{N}_i}+3\gamma_{22_i}+2\gamma^{eq(5)}_{t_i}+2\gamma^{eq(6)}_{t_i}+2\gamma^{eq(7)}_{t_i}+\gamma^{eq(8)}_{t_i}+2\gamma^{eq(9)}_{t_i} \right)  \nonumber\\
  &- \left( \frac{Y_{\Delta \ell}}{Y^{eq}_{\ell}}+\frac{ Y_{\Delta \widetilde{\ell}}}{Y^{eq}_{\widetilde{\ell}}}+\frac{Y_{\Delta H_u}}{Y^{eq}_{H_u}}+\frac{Y_{\Delta \tilde{H}_u}}{Y^{eq}_{\widetilde{H}_u}}\right)  \left( \gamma^{eq}_{\widetilde{N}_i\alpha}+\frac{1}{2}\gamma^{eq}_{N_i\alpha}\right)\nonumber\\
  &-\left( \frac{ Y_{\Delta \widetilde{\ell}}}{Y^{eq}_{\widetilde{\ell}}}+2\frac{Y_{\Delta \tilde{H}_u}}{Y^{eq}_{\widetilde{H}_u}}-\frac{Y_{\Delta H_u}}{Y^{eq}_{H_u}} \right)\left( 2+\frac{1}{2}\frac{Y_{+_i}}{Y^{eq}_{\widetilde{N^c}_i}} \right)\gamma^{eq}_{22_i\alpha}\nonumber\\
  &-\frac{Y_{\Delta \ell}}{Y^{eq}_{\ell}}\left( \frac{Y_{N_i}}{Y^{eq}_{N_i}}\gamma^{eq(3)}_{t_i\alpha}+2\gamma^{eq(4)}_{t_i\alpha}+\frac{Y_{+_i}}{Y^{eq}_{\widetilde{N^c}_i}}\gamma^{eq(5)}_{t_i\alpha}+2\gamma^{eq(6)}_{t_i\alpha}+2\gamma^{eq(7)}_{t_i\alpha} \right)\nonumber\\
  &-\frac{ Y_{\Delta \widetilde{\ell}}}{Y^{eq}_{\widetilde{\ell}}}\left(2\frac{Y_{N_i}}{Y^{eq}_{N_i}}\gamma^{eq(0)}_{t_i\alpha}+2\gamma^{eq(1)}_{t_i\alpha}+2\gamma^{eq(2)}_{t_i\alpha}+\frac{1}{2}\frac{Y_{+_i}}{Y^{eq}_{\widetilde{N}^c_i}}\gamma^{eq(8)}_{t_i\alpha}+2\gamma^{eq(9)}_{t_i\alpha}\right)\nonumber\\
  &-\frac{Y_{\Delta H_u}}{Y^{eq}_{H_u}}\left\{ \frac{1}{2}\left(\frac{Y_{+_i}}{Y^{eq}_{\widetilde{N^c}_i}}-2\right)\left( -\gamma^{eq(6)}_{t_i}+\gamma^{eq(7)}_{t_i} \right)+\left(1+\frac{1}{2}\frac{Y_{+_i}}{Y^{eq}_{\widetilde{N}^c_i}}\right)\gamma^{eq(9)}_{t_i}+\gamma^{eq(8)}_{t_i} \right\}\nonumber\\
  &-\frac{Y_{\Delta H_u}}{Y^{eq}_{H_u}}\left\{\left(\frac{Y_{N_i}}{Y^{eq}_{N_i}}-1\right)\gamma^{eq(1)}_{t_i\alpha}-\frac{Y_{N_i}}{Y^{eq}_{N_i}}\gamma^{eq(2)}_{t_i\alpha}+\gamma^{eq(3)}_{t_i\alpha}+\left(\frac{Y_{N_i}}{Y^{eq}_{N_i}}+1\right)\gamma^{eq(4)}_{t_i\alpha}\right\}\nonumber\\
  &-2\frac{Y_{\Delta \tilde{H}_u}}{Y^{eq}_{\widetilde{H}_u}}\left( \gamma^{eq(0)}_{t_i}+\gamma^{eq(1)}_{t_i\alpha}+\frac{Y_{N_i}}{Y^{eq}_{N_i}}\gamma^{eq(2)}_{t_i\alpha}+\gamma^{eq(5)}_{t_i\alpha}+\frac{1}{2}\frac{Y_{+_i}}{Y^{eq}_{\widetilde{N^c}_i}}\gamma^{eq(6)}_{t_i\alpha}+\gamma^{eq(7)}_{t_i\alpha} \right)\nonumber\\
  &\left.+\frac{Y_{-i}}{Y^{eq}_{\widetilde{N^c}_i}}\left( 2\gamma^{eq(5)}_{t_i\alpha}+2\gamma^{eq(6)}_{t_i\alpha}+2\gamma^{eq(7)}_{t_i\alpha}-\gamma^{eq(8)}_{t_i\alpha}-2\gamma^{eq(9)}_{t_i\alpha} \right)\right],
\end{align}
where,
\begin{align}
  Y_{\Delta b,\Delta f}:=&\frac{1}{g_{b,f}}\frac{\Delta n_{b,f}}{s} \hspace{20pt}(\text{boson,fermion}),\\
  Y_{+ i}:=&Y_{\widetilde{N}_i}+Y_{{\widetilde{N}^{\ast}_i}}\hspace{20pt}(i=1,2),\\
  Y_{- i}:=&Y_{\widetilde{N}_i}-Y_{{\widetilde{N}^{\ast}_i}}\hspace{20pt}(i=1,2),\\
  \frac{Y_{\Delta \ell_\alpha}}{Y^{eq}_{\ell_\alpha}}:=&\frac{Y_{{\ell_L}_\alpha}-Y_{{\overline{\ell}_L}_\alpha}}{Y^{eq}_{\ell_\alpha}}= \frac{ Y_{\Delta \widetilde{\ell}_\alpha}}{Y^{eq}_{\widetilde{\ell}_\alpha}}= -\sum_\beta C^L_{\alpha\beta}\frac{Y_{\Delta_\beta }}{Y^{eq}_{\ell_\alpha}}, \\
  \frac{Y_{\Delta H_u}}{Y^{eq}_{H_u}}:=&\frac{Y_{H_u}-Y_{\overline{H_u}}}{Y^{eq}_{H_u}}= \frac{Y_{\Delta \tilde{H}_u}}{Y^{eq}_{\widetilde{H}_u}}= -\sum_\beta C^H_{\beta}\frac{Y_{\Delta_\beta}}{Y^{eq}_{H_u}},
\end{align}
and equilibrium density
\begin{align}
  Y^{eq}_\ell= \frac{1}{2}Y_{\widetilde{\ell}}^{eq}=\frac{1}{2}Y^{eq}_{H_u}=Y^{eq}_{\widetilde{H}_u}=\frac{15}{8\pi^2g_\ast}.
\end{align}

When the effects of interactions are incorporated into the Boltzmann equations, thermally averaged reaction rates are used.
In particular, the reaction densities for the decay of RH (s)neutrinos and for scatterings involving (s)tops are given by 
\begin{align*}
  \gamma^{eq}_{N_i\alpha}:=& \gamma\left( N_i\leftrightarrow \widetilde{\ell}_\alpha\widetilde{H}_u \right)+\gamma\left( N_i\leftrightarrow \widetilde{\ell}^\ast_\alpha\overline{\widetilde{H}_u} \right)+\gamma\left(N_i\leftrightarrow \ell_\alpha H_u\right)+\gamma\left(N_i\leftrightarrow \overline{\ell}_\alpha H_u^\ast\right),\\
  \gamma^{eq}_{\widetilde{N}_i\alpha}:=&\gamma\left( \widetilde{N}_i\leftrightarrow \widetilde{\ell}_\alpha H_u \right)+\gamma\left( \widetilde{N}_i\leftrightarrow \overline{\ell}_\alpha\overline{\widetilde{H}_u} \right)+\gamma\left(\widetilde{N}_i^\ast\leftrightarrow \widetilde{\ell}_\alpha^\ast H_u^\ast\right)+\gamma\left(\widetilde{N}_i\leftrightarrow \ell_\alpha \widetilde{H}_u\right),\\
  \gamma_{22_i\alpha}:=&\gamma\left( \widetilde{N}_i\widetilde{\ell}_\alpha\leftrightarrow \widetilde{Q}\widetilde{u}^\ast \right)=\gamma\left( \widetilde{N}_i\widetilde{Q}^\ast\leftrightarrow \widetilde{\ell}_\alpha^\ast\widetilde{u}^\ast \right)=\gamma\left( \widetilde{N}_i\widetilde{u}\leftrightarrow \widetilde{\ell}_\alpha^\ast\widetilde{Q}\right),\\
  \gamma^{eq(0)}_{t_i\alpha}:=&\gamma\left( \widetilde{N}_i \widetilde{\ell}_\alpha\leftrightarrow Q\widetilde{u}^\ast \right)=\gamma\left( \widetilde{N}_i \widetilde{\ell}_\alpha\leftrightarrow \widetilde{Q}\overline{u}\right),\\
  \gamma^{eq(1)}_{t_i\alpha}:=&\gamma\left( \widetilde{N}_i \overline{Q}\leftrightarrow \widetilde{\ell}_\alpha^\ast\widetilde{u}^\ast \right)=\gamma\left( \widetilde{N}_i u\leftrightarrow \widetilde{\ell}_\alpha^\ast\widetilde{Q}\right),\\
  \gamma^{eq(2)}_{t_i\alpha}:=&\gamma\left( N_i\widetilde{u}\leftrightarrow \widetilde{\ell}_\alpha^\ast Q \right)=\gamma\left( N_i\widetilde{Q}^\ast\leftrightarrow \widetilde{\ell}_\alpha^\ast \overline{u} \right),\\
  \gamma^{eq(3)}_{t_i\alpha}:=&\gamma\left( N_i\ell_\alpha\leftrightarrow Q\overline{u} \right),\\
  \gamma^{eq(4)}_{t_i\alpha}:=&\gamma\left( N_iu\leftrightarrow \overline{\ell}_\alpha Q \right)=\gamma\left( N_i\overline{Q}\leftrightarrow \overline{\ell}_\alpha \overline{u} \right),\\
  \gamma^{eq(5)}_{t_i\alpha}:=&\gamma\left( \widetilde{N}_i\ell_\alpha\leftrightarrow Q\widetilde{u}^\ast \right)=\gamma\left( \widetilde{N}_i\ell_\alpha\leftrightarrow \widetilde{Q}\overline{u} \right),\\
  \gamma^{eq(6)}_{t_i\alpha}:=&\gamma\left( \widetilde{N}_i \widetilde{u}\leftrightarrow \overline{\ell}_\alpha Q \right)=\gamma\left( \widetilde{N}_i \widetilde{Q}^\ast\leftrightarrow \overline{\ell}_\alpha \overline{u} \right)\\
  \gamma^{eq(7)}_{t_i\alpha}:=&\gamma\left( \widetilde{N}_i \overline{Q} \leftrightarrow \overline{\ell}_\alpha\widetilde{u}^\ast \right)=\gamma\left( \widetilde{N}_i u\leftrightarrow \overline{\ell}_\alpha \widetilde{Q} \right),\\
  \gamma^{eq(8)}_{t_i\alpha}:=&\gamma\left( \widetilde{N}\widetilde{\ell}_\alpha^\ast \leftrightarrow \overline{Q}u \right),\\
  \gamma^{eq(9)}_{t_i\alpha}:=&\gamma\left( \widetilde{N}_i Q \leftrightarrow \widetilde{\ell}_\alpha u \right)=\gamma\left( \widetilde{N}_i\overline{u} \leftrightarrow \widetilde{\ell}_\alpha \overline{Q} \right),
\end{align*}
and
\begin{gather*}
  \gamma^{eq}_{N_i}:=\sum_\alpha \gamma^{eq}_{N_i\alpha},\hspace{10pt}
  \gamma^{eq}_{\widetilde{N}_i}:=\sum_\alpha \gamma^{eq}_{\widetilde{N}_i\alpha},\hspace{10pt}
  \gamma^{eq}_{22_i}:=\sum_\alpha \gamma^{eq}_{22_i\alpha},\\
  \gamma^{eq(n)}_{t_i}:=\sum_\alpha \gamma^{eq(n)}_{t_i\alpha}\,(n=1,2,3,\ldots,9).
\end{gather*}
The details of the above reaction densities are given in \cite{Plumacher:1998ex}.

\newpage
\section{
Sign-flipped Dirac neutrino Yukawa and flavored CP asymmetries
}
In this appendix, we examine the consequences of flipping the signs of individual entries in the Dirac neutrino Yukawa matrix. 
The purpose is to show how to classify the sign-flipped patterns of Dirac neutrino Yukawa couplings so that each group gives the same prediction of the Baryon asymmetry. 

First, the flavored decay parameters do not change at all by changing the sign of the Dirac neutrino Yukawa couplings. Therefore, we focus on the flavored CP asymmetries. 

To begin, we introduce a notation for denoting sign-flip operations of Dirac neutrino Yukawa components. 
Specifically, we write $[(y_{\nu_D})_{\alpha i},\, (y_{\nu_D})_{\beta j},\, \ldots]$ to indicate a procedure in which the signs of the listed Yukawa components are flipped, i.e., $(y_{\nu_D})_{\alpha i} \rightarrow -\,(y_{\nu_D})_{\alpha i}, \,(y_{\nu_D})_{\beta j} \rightarrow -\,(y_{\nu_D})_{\beta j}, \,\ldots$

Among these operations, some of them leave the flavored CP asymmetries unchanged. 
For example, the operations
$[(y_{\nu_D})_{e i},\, (y_{\nu_D})_{\mu i},\, (y_{\nu_D})_{\tau i}],$ $ \, (i = 1,2,3)$ do not change the flavored CP asymmetries.
As an illustration, the asymmetry $\epsilon_{e1}$ under the flip $[(y_{\nu_D})_{e1},\, (y_{\nu_D})_{\mu1},\, (y_{\nu_D})_{\tau1}]$ becomes
\begin{align}
  \epsilon_{e1}&= -\frac{3}{4\pi}\frac{1}{(y_{\nu_D}^\dag y_{\nu_D})_{11}}\left[ \Im\left[\left(y_{\nu_D}\right)_{e1}^\ast\left(y_{\nu_D}\right)_{e2}(y_{\nu_D}^\dag y_{\nu_D})_{12}\right]f\left(\frac{M_2^2}{M_1^2}\right)\right.\nonumber\\
  &\hspace{100pt}\left.+\Im\left[\left(y_{\nu_D}\right)_{e1}^\ast\left(y_{\nu_D}\right)_{e3}(y_{\nu_D}^\dag y_{\nu_D})_{13}\right]f\left(\frac{M_3^2}{M_1^2}\right) \right]\\
  \rightarrow& -\frac{3}{4\pi}\frac{1}{(y_{\nu_D}^\dag y_{\nu_D})_{11}}\left[ \Im\left[\left(-y_{\nu_D}\right)_{e1}^\ast\left(y_{\nu_D}\right)_{e2}(-y_{\nu_D}^\dag y_{\nu_D})_{12}\right]f\left(\frac{M_2^2}{M_1^2}\right)\right.\nonumber\\
  &\hspace{100pt}\left.+\Im\left[\left(-y_{\nu_D}\right)_{e1}^\ast\left(y_{\nu_D}\right)_{e3}(-y_{\nu_D}^\dag y_{\nu_D})_{13}\right]f\left(\frac{M_3^2}{M_1^2}\right) \right]\nonumber\\
  &=-\frac{3}{4\pi}\frac{1}{(y_{\nu_D}^\dag y_{\nu_D})_{11}}\left[ \Im\left[\left(y_{\nu_D}\right)_{e1}^\ast\left(y_{\nu_D}\right)_{e2}(y_{\nu_D}^\dag y_{\nu_D})_{12}\right]f\left(\frac{M_2^2}{M_1^2}\right)\right.\nonumber\\
  &\hspace{100pt}+\left.\Im\left[\left(y_{\nu_D}\right)_{e1}^\ast\left(y_{\nu_D}\right)_{e3}(y_{\nu_D}^\dag y_{\nu_D})_{13}\right]f\left(\frac{M_3^2}{M_1^2}\right) \right]=\epsilon_{e1}.
  \label{samei}
\end{align}
It is easily shown that all the other $\epsilon_{\alpha i}$ are invariant under this type of flip.

Similarly, it is shown that the operations $[(y_{\nu_D})_{\alpha 1},\, (y_{\nu_D})_{\alpha 2},\, (y_{\nu_D})_{\alpha 3}],$ ($\alpha = e, \mu, \tau)$ do not change the flavored CP asymmetries.

By exploiting the above two types of operations, flipping an entire column or an entire row,
it follows that any arbitrary sign-flip configuration involving three or more entries can always be reduced to a configuration involving only one or two entries. 
As a result, the full set of possible sign-flip patterns can be classified into 15 groups as in Table \ref{tab:grouped_yukawa}.

\begin{table}
  \centering
  \caption{15 Groups among all operations for flipping the signs of Dirac neutrino Yukawa couplings. The Yukawa models which are given by some sign flip operations in each group give the same prediction on the Baryon asymmetry. 45 operations which involve only one or two entries are explicitly shown below.
  }
  \begin{tabular}{c|l}
    \hline
    group & flipped components \\
    \hline
    \hline
    $1.$ & $[\left(y_{\nu_D}\right)_{e1}]$, \hspace{2pt} $[\left({y_{\nu_D}}\right)_{e2},\left({y_{\nu_D}}\right)_{e3}]$, \hspace{2pt} $[\left({y_{\nu_D}}\right)_{\mu1},\left({y_{\nu_D}}\right)_{\tau1}]$, ...\\
    $2.$ & $[\left(y_{\nu_D}\right)_{\mu1}]$, \hspace{2pt} $[\left({y_{\nu_D}}\right)_{\mu2},\left({y_{\nu_D}}\right)_{\mu3}]$, \hspace{2pt} $[\left({y_{\nu_D}}\right)_{e1},\left({y_{\nu_D}}\right)_{\tau1}]$, ...\\
    $3.$ & $[\left(y_{\nu_D}\right)_{\tau1}]$, \hspace{2pt} $[\left({y_{\nu_D}}\right)_{\tau2},\left({y_{\nu_D}}\right)_{\tau3}]$\hspace{2pt} $[\left({y_{\nu_D}}\right)_{e1},\left({y_{\nu_D}}\right)_{\mu1}]$, ...\\
    $4.$ & $[\left(y_{\nu_D}\right)_{e2}]$, \hspace{2pt} $[\left({y_{\nu_D}}\right)_{e1},\left({y_{\nu_D}}\right)_{e3}]$, \hspace{2pt} $[\left({y_{\nu_D}}\right)_{\mu2},\left({y_{\nu_D}}\right)_{\tau2}]$, ...\\
    $5.$ & $[\left(y_{\nu_D}\right)_{\mu2}]$, \hspace{2pt} $[\left({y_{\nu_D}}\right)_{\mu1},\left({y_{\nu_D}}\right)_{\mu3}]$, \hspace{2pt} $[\left({y_{\nu_D}}\right)_{e2},\left({y_{\nu_D}}\right)_{\tau2}]$, ...\\
    $6.$ & $[\left(y_{\nu_D}\right)_{\tau2}]$, \hspace{2pt} $[\left({y_{\nu_D}}\right)_{\tau1},\left({y_{\nu_D}}\right)_{\tau3}]$, \hspace{2pt} $[\left({y_{\nu_D}}\right)_{e2},\left({y_{\nu_D}}\right)_{\mu2}]$, ...\\
    $7.$ & $[\left(y_{\nu_D}\right)_{e3}]$, \hspace{2pt} $[\left({y_{\nu_D}}\right)_{e1},\left({y_{\nu_D}}\right)_{e2}]$, \hspace{2pt} $[\left({y_{\nu_D}}\right)_{\mu3},\left({y_{\nu_D}}\right)_{\tau3}]$, ...\\
    $8.$ & $[\left(y_{\nu_D}\right)_{\mu3}]$, \hspace{2pt} $[\left({y_{\nu_D}}\right)_{\mu1},\left({y_{\nu_D}}\right)_{\mu2}]$, \hspace{2pt} $[\left({y_{\nu_D}}\right)_{e3},\left({y_{\nu_D}}\right)_{\tau3}]$, ...\\
    $9.$ & $[\left(y_{\nu_D}\right)_{\tau3}]$, \hspace{2pt} $[\left({y_{\nu_D}}\right)_{\tau1},\left({y_{\nu_D}}\right)_{\tau2}]$, \hspace{2pt} $[\left({y_{\nu_D}}\right)_{e3},\left({y_{\nu_D}}\right)_{\mu3}]$, ...\\
    $10.$& $[\left({y_{\nu_D}}\right)_{e1},\left({y_{\nu_D}}\right)_{\mu2}]$, \hspace{2pt} $[\left(y_{\nu_D}\right)_{e3},\left(y_{\nu_D}\right)_{\tau2}]$, \hspace{2pt} $[\left(y_{\nu_D}\right)_{\mu3},\left(y_{\nu_D}\right)_{\tau1}]$, ...\\
    $11.$& $[\left({y_{\nu_D}}\right)_{e1},\left({y_{\nu_D}}\right)_{\tau2}]$, \hspace{2pt} $[\left(y_{\nu_D}\right)_{e3},\left(y_{\nu_D}\right)_{\mu2}]$, \hspace{2pt} $[\left(y_{\nu_D}\right)_{\mu1},\left(y_{\nu_D}\right)_{\tau3}]$, ...\\
    $12.$& $[\left({y_{\nu_D}}\right)_{e1},\left({y_{\nu_D}}\right)_{\mu3}]$, \hspace{2pt} $[\left(y_{\nu_D}\right)_{e2},\left(y_{\nu_D}\right)_{\tau3}]$, \hspace{2pt} $[\left(y_{\nu_D}\right)_{\mu2},\left(y_{\nu_D}\right)_{\tau1}]$, ...\\
    $13.$& $[\left({y_{\nu_D}}\right)_{e1},\left({y_{\nu_D}}\right)_{\tau3}]$, \hspace{2pt} $[\left(y_{\nu_D}\right)_{e2},\left(y_{\nu_D}\right)_{\mu3}]$, \hspace{2pt} $[\left(y_{\nu_D}\right)_{\mu1},\left(y_{\nu_D}\right)_{\tau2}]$, ...\\
    $14.$& $[\left({y_{\nu_D}}\right)_{\mu1},\left({y_{\nu_D}}\right)_{e2}]$, \hspace{2pt} $[\left(y_{\nu_D}\right)_{e3},\left(y_{\nu_D}\right)_{\tau1}]$, \hspace{2pt} $[\left(y_{\nu_D}\right)_{\mu3},\left(y_{\nu_D}\right)_{\tau2}]$, ...\\
    $15.$& $[\left({y_{\nu_D}}\right)_{\mu1},\left({y_{\nu_D}}\right)_{e3}]$, \hspace{2pt} $[\left(y_{\nu_D}\right)_{e2},\left(y_{\nu_D}\right)_{\tau1}]$, \hspace{2pt} $[\left(y_{\nu_D}\right)_{\mu2},\left(y_{\nu_D}\right)_{\tau3}]$, ...\\
    \hline
  \end{tabular}
  \label{tab:grouped_yukawa}
\end{table}

\newpage
\clearpage
\newpage
\bibliographystyle{JHEP}
\bibliography{bibliography}
\end{document}